# Optical, electronic and structural properties of ScAlMgO$_4$

*Tomasz Stefaniuk[a]\*, Jan Suffczyński[a], Małgorzata Wierzbowska[b], Jarosław Z. Domagała[c], Jarosław Kisielewski[d], Andrzej Kłos[d], Alexander Korneluk[a], Henryk Teisseyre[b,c]*

[a] Faculty of Physics, University of Warsaw, 5 Pasteura St., 02-093 Warsaw, Poland

[b] Institute of High Pressure Physics Polish Academy of Sciences, 29 Sokołowska St., 01-142 Warsaw, Poland

[c] Institute of Physics, Polish Academy of Sciences, 32/46 Al. Lotników, 02-668 Warsaw, Poland

[d] Łukasiewicz Research Network - Institute of Microelectronics and Photonics, Wólczyńska 133, 01-919 Warsaw, Poland

ABSTRACT

Magnesium aluminate scandium oxide (ScAlMgO$_4$) is a promising lattice-matched substrate material for GaN- and ZnO-based optoelectronic devices. Yet, despite its clear advantages over substrates commonly used in heteroepitaxial growth, several fundamental properties of ScAlMgO$_4$ remain unsettled. Here, we provide a comprehensive picture of its optical, electronic and structural properties by studying ScAlMgO$_4$ single crystals grown by the Czochralski method. We use variable angle spectroscopic ellipsometry to determine complex in-plane and out-of-plane refractive indices in the range from 193 to 1690 nm. An oscillator-based model provides a phenomenological description of the ellipsometric spectra with excellent agreement over the entire range of wavelengths. For convenience, we supply the reader also with Cauchy formulas describing the real part of the anisotropic refractive index for wavelengths above 400 nm. *Ab initio* many-body perturbation theory modeling provides information about the electronic structure of ScAlMgO$_4$, and successfully validated experimentally obtained refractive index values. Simulations also show exciton binding energy as large as a few hundred of meV, indicating ScAlMgO$_4$ as a promising material for implementation in low-threshold, deep-UV lasing devices operating at room temperature. X-ray diffraction measurements confirm lattice constants of ScAlMgO$_4$ previously reported, but in addition, reveal that dominant crystallographic planes (001) are mutually inclined by about 0.009°. In view of our work, ScAlMgO$_4$ is a highly transparent, low refractive index, birefringent material similar to a sapphire, but with a much more favorable lattice constant and simpler processing.

1. **INTRODUCTION**

In recent years magnesium aluminate scandium oxide (ScAlMgO$_4$ or SCAM) has been suggested as an optimal substrate for the epitaxial growth of GaN- and ZnO-based materials[1-5]. This is due to the advantages of SCAM over other materials typically used for that purpose. Firstly, SCAM's lattice mismatch between SCAM and GaN, and between SCAM and ZnO is much smaller than that between SCAM and sapphire, widely used in current semiconductor optoelectronic technology as a substrate material[6]. Therefore, the growth on the SCAM substrate reduces the dislocation density, which improves the electrical and optical properties of a device. Secondly, SCAM has a closely matched thermal expansion coefficient along the *a*-axis[1,6] with that of GaN and ZnO, enabling a decrease of the residual strain in GaN or ZnO epitaxial layers. Thirdly, SCAM is easily cleaved along the *c*-plane. This, in turn, allows atomically flat *c*-plane substrates to be obtained without polishing[7,8], leading to a reduction of device production costs. Fourthly, $ScAlMgO_4$ can be grown in the form of large crystals. In particular, using the Czochralski method, wafers of more than 2 inches in diameter can be obtained[7], further reducing the substrate cost. Finally, SCAM is physicochemical stable[8] and has high optical transparency in the visible spectral range, which is essential for many real-world applications. These attributes make SCAM a good candidate for usage in semiconductor-based electronic devices.

$ScAlMgO_4$ possesses a crystallographic structure of $YbFe_2O_4$ type (space group $R\bar{3}m$)[5]. Its structure exhibits a crystalline anisotropy because it consists of an alternating stack of wurtzite $(Mg, Al)O_x(0001)$ and rock-salt $ScO_y(111)$ layers[3]. The experimentally established lattice parameters are $a = b = 3.245 \pm 0.005$ Å and $c = 25.14 \pm 0.04$ Å (see Table S1 and references therein [1,3,5,9,10]), which ensures a lattice mismatch between $ScAlMgO_4$ and GaN as low as 1.7%, and only 0.15% between SCAM and ZnO. The first attempt to use SCAM as a substrate was made by Hellman[1] *et al.*, who successfully exploited the nitrogen-plasma molecular beam epitaxy (MBE) to grow the GaN layer on $ScAlMgO_4$. Later, it was demonstrated that nitride layers on the SCAM substrates could also be grown by metalorganic vapor phase epitaxy[7]. Recently, a halide vapor phase epitaxy growth of a 320-μm-thick GaN layer[11], with threading dislocation density as low as $2.4 \times 10^7$ cm$^{-2}$, and fabrication of InGaN-based visible light-emitting diodes (LEDs) on the $ScAlMgO_4$ substrate were accomplished[2]. In parallel, reports on the successful growth of the ZnO layers on ScAlMgO$_4$ substrates have started to appear[12,13]. In particular, Ohtomo *et al.* employed the laser MBE technique[3], reaching an extremely smooth surface with atomically flat terraces and roughness at the level of 0.26 nm. The same group reported optically pumped laser emission from ZnO/Zn$_{1-x}$Mg$_x$O superlattice[14] on the SCAM substrate with a very low threshold power density of 11 kW/cm$^2$, comparable with the lasing thresholds in structures where the light-matter interaction is increased thanks to the incorporation of the semiconductor quantum wells to an optical microcavity[15,16].

Successful exploitation of ScAlMgO$_4$ in photonic applications requires precise knowledge of its optical, electronic and structural properties. Only then it is possible to design and optimize the performance of the optical systems built on the SCAM substrates. In this work, we employ optical and X-ray diffraction (XRD) spectroscopy, as well as perform theoretical calculations to establish the properties of SCAM. Variable angle spectroscopic ellipsometry has enabled us to determine the complex dielectric permittivity of ScAlMgO$_4$ single crystal samples. We establish the refractive index value of this material, reveal that it exhibits optical anisotropy (birefringence), and estimate the optical losses. The density functional theory (DFT) calculations performed in the frame of *ab initio* many-body perturbation theory (*ab initio* MBPT) provide electronic structure and a dielectric function of ScAlMgO$_4$ consistent with the results of the ellipsometric experiment. The theoretical calculations point towards a high, sub-eV exciton binding energy in this material. The XRD measurements bring lattice constants of ScAlMgO$_4$ consistent with previous reports, but in addition, they indicate that dominant crystallographic planes (001) are mutually inclined by about 0.009°. We complement our work with characterization by means of Scanning Electron Microscopy and Raman spectroscopy to bring a comprehensive description of the ScAlMgO$_4$ properties.

## 2. SAMPLES AND METHODS

**2.1 Sample Fabrication.** ScAlMgO$_4$ single crystals, used in our studies, are grown by the Czochralski method using the Oxypuller system produced by Cyberstar. The heating system is based on a Hüttinger generator. The growth processes are carried out in the iridium crucible with a diameter of 50 mm and a height of 50 mm, embedded in a Zircar zirconia grog. Pure nitrogen is used as a protective atmosphere. The ScAlMgO$_4$ crystals are grown by chemical reaction with stoichiometric amounts of 4.5 N Magnesium Oxide (MgO), aluminum oxide (Al$_2$O$_3$), and scandium oxide (Sc$_2$O$_3$), with a pulling rate of 1.2– 2 mm/h and rotation rate from 6 to 20 rpm. With the described approach, good optical quality single crystals up to 50 mm in length and 20 mm in diameter are obtained (see Figure 1(a)). The layered nature of the crystals is confirmed by scanning electron microscopy (SEM), as shown in Figure 1(b). The stoichiometric properties of the very same crystals have been measured and reported previously in the work of Wierzbicka[17] *et al.*, where Energy-dispersive X-ray spectroscopy was performed to detect possible composition heterogeneity. Microanalysis revealed no other elements except those included in ScAlMgO$_4$ (Fig. 15 in the mentioned reference). Distribution maps made in different areas (40 μm x 60 μm) showed no spatial changes in chemical composition in the tested crystals.

**2.2 X-ray Measurements.** The X-ray measurements are performed using a Philips X'Pert MRD high-resolution X-ray diffractometer, working with copper radiation (Cu$_{K\alpha1}$, $\lambda$ = 1.5406 Å) and filtered by a 4xGe(220) asymmetric monochromator. Both the double- and triple-crystal modes of the device are employed to collect the curves. In the double mode, we performed so-called rocking curve (RC) scans. In such a scan, the detector does not move, and it is fully exposed, collecting intensity as a function of the *omega* being the angle of the X-ray beam incidence on the crystal surface. The setup works in the triple-crystal mode (TA), where the beam reflected by the sample passes through the analyzer crystal (Ge[220] – 3-bounce channel cut) before reaching the detector. This diffractometer geometry allows us to separate the effects resulting from the so-called mosaicity (flatness defects caused by the plane bending or linear defects) or the presence of grains, from the dispersion of interplanar distances. The flatness is tested by collecting the intensity curves respectively as a function of the incidence angle *omega$_{(TA)}$* (*omega* in the triple-crystal mode). The interplanar spread is examined by detecting the intensity curves as a function of interconnected detector motion (*2theta* angle) with a change in the *omega*. In the latter case, the angle of the detector changes 2 times faster than the *omega* angle.

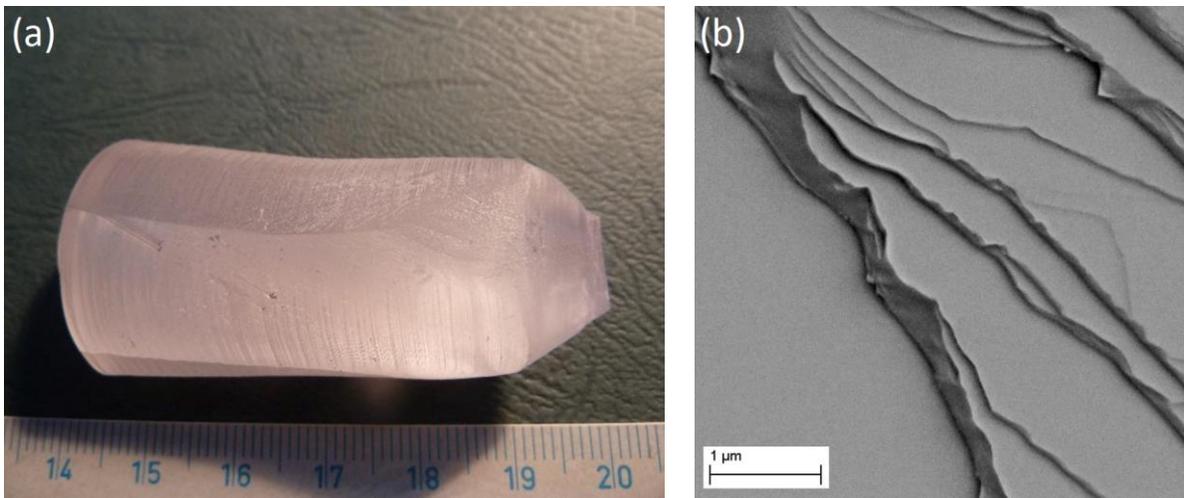

**Figure 1.** (a) A photograph of a ScAlMgO$_4$ single crystal grown by the Czochralski method (the scale is in cm). (b) The image of the step-terrace structure of the cleaved sample taken using scanning electron microscopy. The height of the steps varies between 10 nm to 300 nm.

**2.3 Raman Spectroscopy.** Raman spectra are recorded in the backscattering geometry with the **k** vector parallel to the crystal's *c* axis (*z*-direction). The signal is excited at 532 nm with a power of 5 mW per 2 μm diameter spot. Measurements are performed at ambient conditions (295 K) or after placing the sample in a pumped helium cryostat at 1.5 K. A Peltier-cooled CCD camera mounted on the output of a 500-mm-long, grating spectrometer (1200 gr/mm) serves as a detector. The spectral resolution of the setup is better than 1 cm$^{-1}$.

**2.4. Spectroscopic Ellipsometry.** The optical properties of SCAM are investigated using variable angle spectroscopic ellipsometry (SE). The technique relies on the measurement of a change of the polarization state of the light reflected from the surface of the sample after incidence from a non-normal direction. In contrast to reflectometry, SE is insensitive to fluctuations in light intensity and, in general, allows the complex refractive index of the material, birefringence, roughness, and thickness of the sample to be determined. Standard ellipsometry, which measures the pair of ellipsometric parameters, namely $\psi(\omega)$ and $\Delta(\omega)$, is typically used only for isotropic samples because the cross-polarization between the *p* and *s* orientations is ignored. The optically anisotropic samples exhibit cross-polarization effects, and the analysis requires a more advanced technique, called generalised ellipsometry in this case[18]. However, the optical properties of the samples that depolarize the light still cannot be quantified[18,19]. Finally, the most general Mueller matrix (MM) ellipsometry is able to describe both the cross-polarization and depolarization effects. This makes the MM ellipsometry a versatile technique for optical characterization of any sample type[20,21]. We implement this technique in the present work.

Our investigations are conducted with an RC2 ellipsometer (manufactured by J.A. Woollam Co.) equipped with dual-rotating compensators placed before and after the sample. All data are collected using a collimated beam with a diameter of 3-4 mm. For our studies, we prepared 3 *c*-plane SCAM lamellas by cleaving the boule with a razor blade. We limit our research to samples with *c*-plane orientation because, due to the layered nature of the material, mechanical polishing would introduce surface defects in the case of other cuts. To increase the sensitivity of our measurement to the anisotropy, samples with the thickness of 0.273 mm (#1) and 0.564 mm (#2) are left as cleaved, i.e. with both interfaces smooth. The third sample with a thickness of 0.65 mm (#3) is mechanically roughened on one side, limiting the responsiveness of the SE measurements to ordinary refractive index values. To obtain complete information about the optical properties in different directions, we measure 16 MM elements in the reflection (samples #1, #2, and #3) and the transmission (samples #1 and #2) over a wide spectral range from 193 to 1690 nm. The MM data are acquired in a transmission mode for a wide angle of incidence ranging from -5° to 50° by 5° and in the reflection-mode MM for an angle of incidence from 55° to 75° by 5°. For brevity, only a few angles are displayed throughout the figures in this work. Additionally, the spectroscopic ellipsometry data are supplemented with the transmission intensity data gathered from 0° to 40° by 10° using an RC2 ellipsometer for the double-side smooth samples. Such a combination reduces the possible correlation between the sample thickness and optical constants, leading to a unique solution. All further data analysis is performed using CompleteEASE software[22]. Additional information regarding the ellipsometric measurements and data analysis procedures are available in Supporting Information.

**2.5. Theoretical Approach.** Prior to simulations of the optical properties, we perform the density functional theory[23] calculations using the plane-wave Quantum ESPRESSO package[24]. The atomic nuclei and core electrons are treated with the pseudopotentials of the norm-conserving Martin-Troulier type. The Perdew-Burke-Ernzerhof parametrization of the exchange-correlation functional[25] is used with the plane-wave energy cut-off of 60 Ry. The Brillouin zone (BZ) is sampled with the 4 × 4 × 1 Monkhorst-Pack mesh[26] in the self-consistent DFT calculations and a dense, 12 × 12 × 2 k-points, sampling for the density of states. The greater density of the BZ sampling along the *a*-axis or *b*-axis than along the c-axis is due to a highly elongated elementary cell of SCAM, as obtained from the XRD. This elementary cell contains 84 atoms, and it is presented in Figure 2.

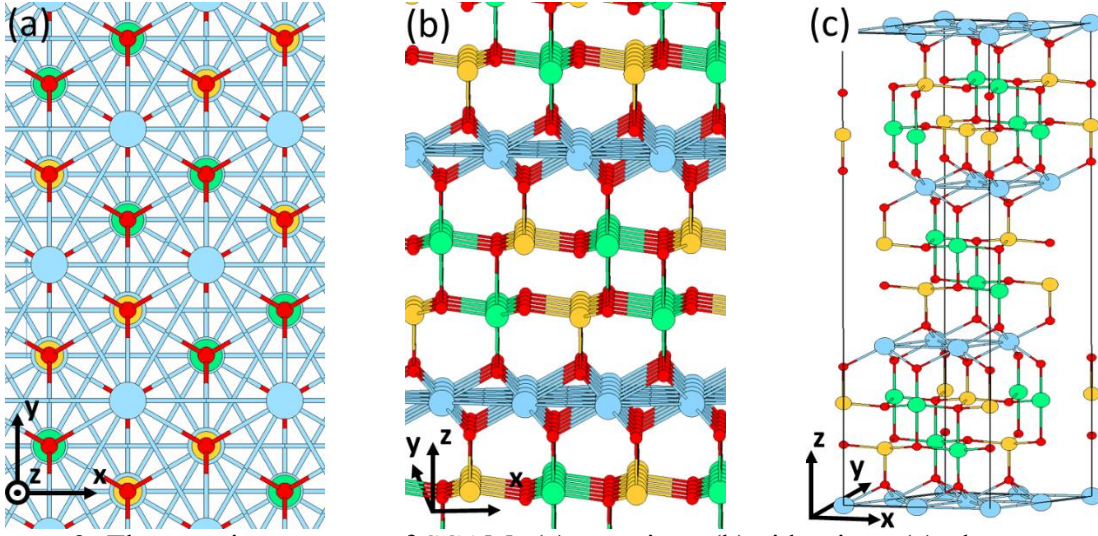

**Figure 2.** The atomic structure of SCAM: (a) top view, (b) side view, (c) elementary cell. The color code for atoms is: Sc – blue, Al – orange, Mg – green, O – red.

Furthermore, the *ab initio* MBPT approach, implemented in the Yambo cod[28], is applied for calculations of the macroscopic inverse dielectric function defined as

$$\varepsilon_M(\omega) \equiv \lim_{\mathbf{q}\to 0} \frac{1}{[\varepsilon(\mathbf{q},\omega)^{-1}]_{\mathbf{G}=0,\mathbf{G}'=0}}. \qquad (1)$$

The quantity $\varepsilon^{-1}_{\mathbf{GG}'}(\mathbf{q},\omega)$ is a matrix in the reciprocal lattice vectors $\mathbf{G}$ and $\mathbf{G}'$

$$\varepsilon^{-1}_{\mathbf{GG}'}(\mathbf{q},\omega) = \delta_{\mathbf{GG}'} + v(\mathbf{q}+\mathbf{G})\chi_{\mathbf{GG}'}(\mathbf{q},\omega), \qquad (2)$$

where $v(\mathbf{q}+\mathbf{G})$ is the bare Coulomb potential. The reciprocal lattice vectors G and G' are set to zero after the matrix inversion. After the matrix inversion, we take only the first matrix element, which is a function of the light frequency (ω) and light polarization $\mathbf{q}$. The quantity $\chi_{\mathbf{GG}'}(\mathbf{q},\omega)$ is the linear response function (LRF), calculated in the random phase approximation (RPA) with the local fields effect as defined in SI. The Eqs. (S1)-(S4) explain how the LRF involves the eigenvalues and eigenvectors, i.e. Bloch functions, obtained from the DFT. The plane-wave energy cut-off of 6 Ry and summation over 384 bands (with 192 being occupied) is enough to converge the LRF.

The real $\varepsilon_1$ and imaginary $\varepsilon_2$ parts of the macroscopic dielectric function $\varepsilon_M$ are embedded in the Kramers-Kronig relations for the refraction ($n$) and extinction ($\kappa$) coefficients, defined as follows:

$$n = \frac{1}{\sqrt{2}}\sqrt{\varepsilon_1 + \sqrt{\varepsilon_1^2 + \varepsilon_2^2}} \tag{3}$$

$$\kappa = \frac{1}{\sqrt{2}}\sqrt{\sqrt{\varepsilon_1^2 + \varepsilon_2^2} - \varepsilon_1} \tag{4}$$

Both the normal and extraordinary components are obtained using the $\varepsilon_M$ calculated with the corresponding polarization of the electric field, which is the **q** vector in Eqs. 1-2.

In addition, the light absorption spectrum is calculated from the Bethe-Salpeter equation[29] (BSE), taking into account the effect of the electron-hole (*e-h*) interactions. The general form of the BSE and its simplification for semiconductors at zero temperature, as well as the BSE dependence on the DFT eigenvectors and eigenvalues, are described in detail in SI. The absorption spectrum is defined via the eigenvalues $\lambda$ and eigenvectors $A^\lambda_{n'n\mathbf{k}} = \langle n'n\mathbf{k}|\lambda\rangle$ of the Hamiltonian described in Supporting Information, Eq. S12, as follows:

$$\varepsilon_M(\omega) \equiv 1 - \lim_{\mathbf{q}\to 0}\frac{8\pi}{|\mathbf{q}|^2 \Omega N_\mathbf{q}}\sum_{nn'\mathbf{k}}\sum_{mm'\mathbf{k}'}\rho^*_{n'n\mathbf{k}}(\mathbf{q},\mathbf{G})\rho_{m'm\mathbf{k}'}(\mathbf{q},\mathbf{G}')\sum_\lambda \frac{A^\lambda_{n'n\mathbf{k}}(A^\lambda_{m'm\mathbf{k}'})^*}{\omega - E_\lambda} \tag{5}$$

The screening matrix elements $\rho(q,G)$ are defined by Eq. S3, $N_q$ is the number of $q$ points in the summation over BZ, and $\Omega$ is the volume of the elementary cell. The indexes *n, n', m,* and *m'* number the bands, and **k**-grid samples of these electronic states in BZ.

Finally, the excitonic binding energy ($E_b$) is defined via the difference between the bandgap value obtained with the GW method ($E_g^{GW}$) and the bandgap value obtained from the low-energy edge of the absorption spectrum calculated with the BSE ($E_g^{BSE}$), as follows

$$E_b = E_g^{GW} - E_g^{BSE} \qquad (6)$$

The GW calculations[30] are performed with the plasmon-pole approximation[31]. The DFT Bloch functions, used as an input, are calculated on the 4 × 4 × 1 grid in BZ.

The plane-wave energy cut-offs for the Coulomb and exchange interactions are set, respectively, to 30 Ry and 6 Ry, and the response block size is set to 6 Ry. Around 600 bands are included in the response, as well as the GW summations. The BSE is solved using a dense 8 × 8 × 1 Monkhorst-Pack mesh to sample the BZ, in order to obtain a convergent absorption spectrum within the Haydock solution scheme. For BSE, 384 bands are summed in the linear response, while 192 bands are occupied. The plane waves up to the energies of 10 Ry for the exchange components and 4 Ry for both the screening and response block size are used. The *e-h* pairs are formed on the 10 occupied and 12 unoccupied bands.

## 3. RESULTS AND DISCUSSION

**3.1 Structural Properties.** The study of the optical properties of a crystal requires reference to its crystallographic quality and quantitative characteristics such as lattice parameters. For this reason, we start with the characterization of the studied layers by X-ray diffraction.

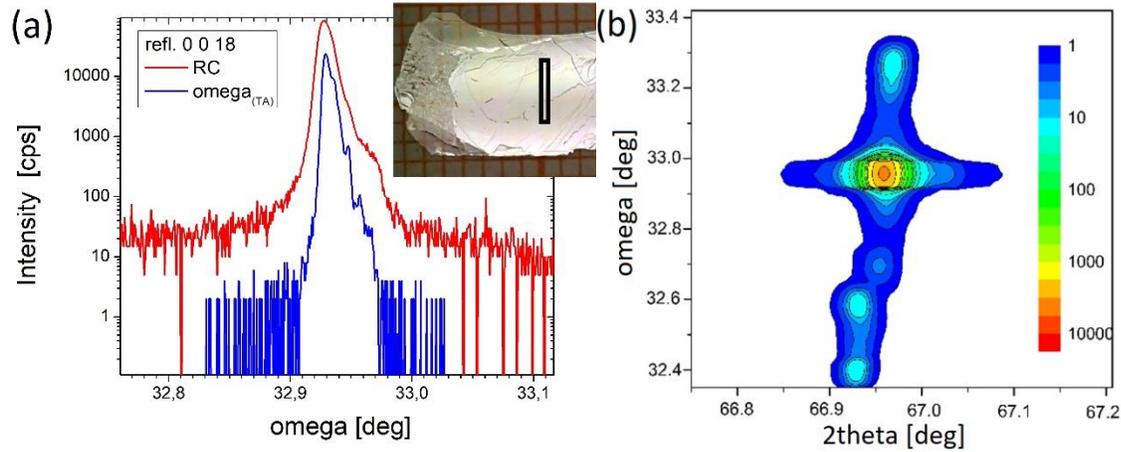

**Figure 3.** (a) Reflection 0 0 18; *omega* scans measured on a ScAlMgO$_4$ layer in double-axis (RC) and triple-axis (*omega$_{(TA)}$*) diffractometer configuration. *2theta/omega* scan coupled to the surface in a wide angular range is provided in the SI. (b) Two-dimensional distribution of the reflected X-ray intensity of the 0 0 18 reflection in the angular axis (241 scans of *2theta/omega* angle with slightly different coupling values of *omega*). The maxima above and below the main node are related to the presence of slight inclination (mutual slope = *Δomega*~0.02° defined by the distance between the maxima) and have different *2theta* coordinates (maximum *Δ2theta*~0.04°). The data are smoothed using a triangular algorithm, developed by Malvern Panalytical, and available in Amass 1.0a software. The intensity color bar is on the logarithmic scale. An illuminated area of 0.3 mm × 3 mm is indicated in the photograph shown in the inset to panel a.

The surface of the tested crystal is parallel to the (001) planes. With the PIXcell semiconductor detector operating in the scanning mode and double crystal detector configuration, we measure a *2theta/omega* scan coupled to the surface in a wide 2theta angular range (from $10°$ to $110°$). We find that only peaks from 9 successive rows of reflections for the (001) plane and nothing else is present in the spectra. This indicates that our material is phase homogeneous. For crystals grown by the Czochralski method, a detailed analysis of the defects investigated by the topography method was done previously by Wierzbicka[17] *et al*.

The shape of the RC curve, especially the full width at half maximum (FWHM), generally informs whether the tested crystal is of good quality. The curve should have one maximum and the FWHM as close as possible to that of an ideal material, e.g. Si, measured with the implemented diffractometer. The crystallographic perfection is tested by means of a 0 0 18 reflection. The RC and *omega$_{(TA)}$* curves, shown in Figure 3(a), are collected with a narrow vertical slit $(1/32°)$ and a horizontal 2 mm mask in the X-ray incident path. They show that within the very small illuminated sample area (0.3 mm × 3 mm), the investigated crystal is almost perfect. The FWHM of the RC and *omega $_{(TA)}$* is $0.012°$ and $0.006°$, respectively. We point out that due to the implemented optics forming the X-ray beam, the maxima in Figure 3(a) have the FWHM only slightly higher than that of the 004 reflection of the ideal silicon $(0.007°)$. However, the *omega$_{(TA)}$* curve, which is very sensitive to any deviations from perfection, shows slight inflections, such as additional weak maxima. This indicates that some of the dominant planes (001) are inclined with respect to the other (001) planes, by about $0.009°$ on average.

We calculate the unit cell parameters for SCAM based on the measurements of *2theta/omega* scans in TA mode of the diffractometer. We start from reflections 0 0 18 and 1 0 1 to determine the *c* parameter. Knowing *c*, we calculate *a*. Because the *2theta/omega* curves for the 0 0 18 reflections, measured at 2 different points on the sample, have slightly different positions of their maxima, we enlarge the illuminated area of the sample to 3 mm × 3 mm, in order to examine as many crystalline planes as possible in one experiment. The obtained two-dimensional map allows us to observe possible deviations in the *c* parameter on a larger spatial scale. The map (see Figure 3(b)) shows a typical picture of a grainy material. Still, the main node on the map has more than 3 orders of magnitude greater intensity than the others. It means that the illuminated area of the sample is dominated by one plate, which proves that we are dealing with a single crystal. Its lattice parameter is $c_{main}$ = 25.1352 Å spread by $\Delta c/c_{main}$ = 5.3×10$^{-4}$. The presence of the additional thin plates with an average angle of 0.02° between each other and slightly different *c* lattice parameters ($\Delta c$~0.0132 Å) is suggested by different *2theta* coordinates for nodes much weaker than the principal one, as seen in Figure 3(b). The lattice parameter *a*, calculated using the $c_{main}$, is equal to 3.246 Å with an uncertainty of ±0.005 Å (a specific measurement geometry and low-value Bragg angle, $2theta_{101}$ = 32.013°, causes such a relatively large uncertainty). The values of the obtained lattice constants are fully consistent with previous reports (see Table S1).

**3.2. Raman-active Vibrational Modes.** The quality of the grown samples was also confirmed via the Raman measurements. In Figure 4 we show the unpolarized Raman spectrum of the ScAlMgO$_4$ sample registered at 295 K and at 1.5 K. The shape and position of the optically active phonon modes at 295 K agree with the previous reports[11]. Upon a decrease of the temperature from 295 K to 1.5 K, the spectrum shape remains practically unchanged. At the same time, the peaks become slightly narrower, and they shift towards the higher energies by around 3 cm$^{-1}$. In particular, the dominating maximum shifts from 418.0 ± 0.9 cm$^{-1}$ at room temperature to 420.9 ± 0.9 cm$^{-1}$ at 1.5 K. In the configuration of the experiment (the light incidence along the *c*-axis), 12 Raman modes are expected for SCAM (6 E$_g$ and 6 A$_{1g}$).

Following Ref. 7, we attribute 2 principal peaks in the spectrum to the $E_g$ modes, as indicated in Figure 3.

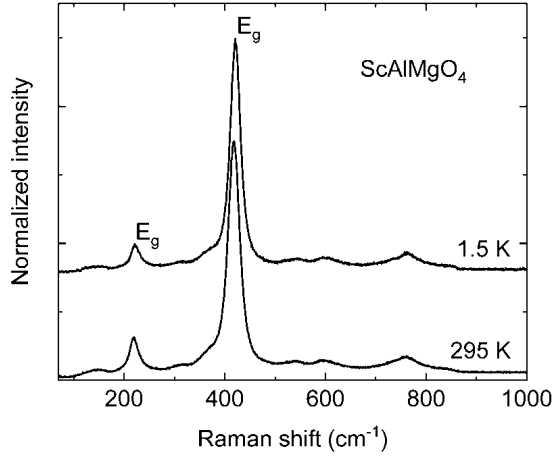

**Figure 4.** The Raman spectrum of $ScAlMgO_4$ single crystal, excited at 532 nm, at temperatures of 1.5 K and 295 K. The spectra are vertically shifted for clarity.

**3.3. Electronic Structure of SCAM.** The DFT modeling specifies that the fundamental energy bandgap ($E_g^{DFT}$) in SCAM equals 3.58 eV. This is, as usual for this method, far below the reported experimental bandgap of 6.29 eV extracted from the transmission curve[32]. However, when the GW correction is taken into account, the bandgap shifts towards the expected value, i.e. 6.32 eV. The GW bandgap is independent of the polarization of the electric field. Because the GW does not include the *e-h* interaction, the above effect should be correlated with the electrical bandgap and not the optical bandgap. In turn, many different results are obtained from the low-energy edge of the absorption spectrum calculated with the BSE and presented in Figure 5(a). These results indicate that the direction of the polarization of the electric field makes a pronounced difference in the spectrum intensity, shape, and the bandgap. The optical bandgaps obtained with the BSE – which includes the *e-h* interactions – are 6.11 eV and 5.77 eV for the planar and perpendicular orientation of the electric field, respectively, and the corresponding exciton binding energies are 0.21 eV and 0.55 eV. The optical bandgap is smaller than the electric (so-called dark) bandgap due to the relaxation of the crystal orbitals in response to the *e-h* interaction.

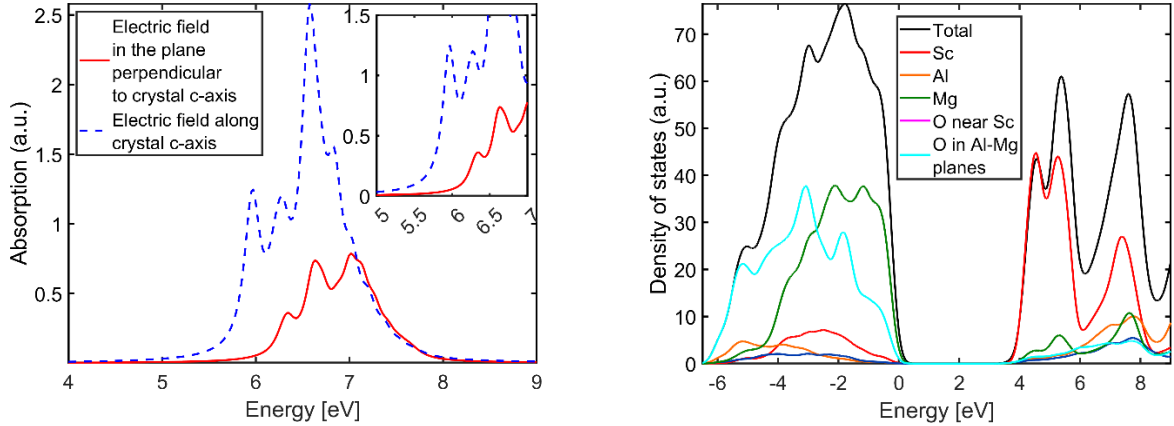

**Figure 5.** Calculated optical and electronic properties of SCAM: (a) the light absorption spectrum obtained with the BSE for 2 polarizations of the electric-field vector: parallel to the crystal *c*-axis and perpendicular to it. The inset shows the low-energy edge of the absorption spectrum. (b) The density of states projected on the atomic shells was obtained with the DFT method.

The large $E_b$ values suggest that $ScAlMgO_4$ would be a good optical material for implementation in lasing devices. Taking into account a large bandgap of the material and high exciton oscillator strength implied by sub-eV binding energy, such devices would exhibit a deep-UV emission, a low threshold and high stability of operation at room temperature conditions.

The conduction band is composed mainly of the Sc-located states, and the valence band is primarily constructed by the oxygen states, with a domination of the oxygens neighboring the Sc over those from the (Al, Mg)O layers (see Figure 5(b)). A comparably large anisotropy of the excitonic properties in layered materials has also been reported for the 2D hybrid halide double perovskites[33] and $MoS_2$ layers embedded in h-BN[34]. It is worth noting that SCAM and both reported materials contain transition metals, like Sc, Ag, and Mo, building strong chemical bonds with the elements of group V or VI in the periodic table. These chemical bonds very strongly localize the *e-h* pairs.

**3.4. Refractive Index and Birefringence.** To determine the dielectric function of SCAM and to support numerical predictions that SCAM exhibits optical anisotropy, we performed an ellipsometry measurement and analyzed the data in the following way: firstly, to identify the axes orientations of a cleaved sample, we focus on the MM data. In Figure 6, we present the MM elements (normalized with respect to the $m_{11}$ element) as a function of wavelength and obtained at multiple angles of reflection for the 0.273-mm-thick sample (transmission ellipsometry data are available in Figure S4).

Almost negligible values of the off-diagonal terms ($0 \pm 0.003$ for all wavelengths) and off-diagonal blocks reveal no cross-polarization between the *p* and *s* states, and they indicate that the sample is either the isotropic or uniaxial crystal with a *c*-plane anisotropy. However, the bottom right terms ($m_{33}$, $m_{34}$, $m_{43}$, $m_{44}$), which are sensitive to retardance, reveal some characteristic oscillations. Because these oscillations are not present in the MM reflection data obtained for the backside roughened sample, we attribute their origin to optical anisotropy. The fringes arise from the birefringence splitting of light that travels through the sample with different velocities, reflects from the backside interface, and returns to the top surface. The amplitude ratio and phase difference of the reflection coefficients of the *p*- and *s*-polarized light varies as a function of the wavelength, allowing us to quantify the birefringence. Because the thickness of the analyzed sample is greater than the coherent length of the RC2 ellipsometer, the simple multiple reflections between the 2 interfaces between the SCAM and air would not create such an interference pattern. Finally, because the sample rotation does not influence the MM elements (see Figure S3), we assume that the *c*-axis of the cleaved SCAM lamellae has high accuracy perpendicular to the sample surface.

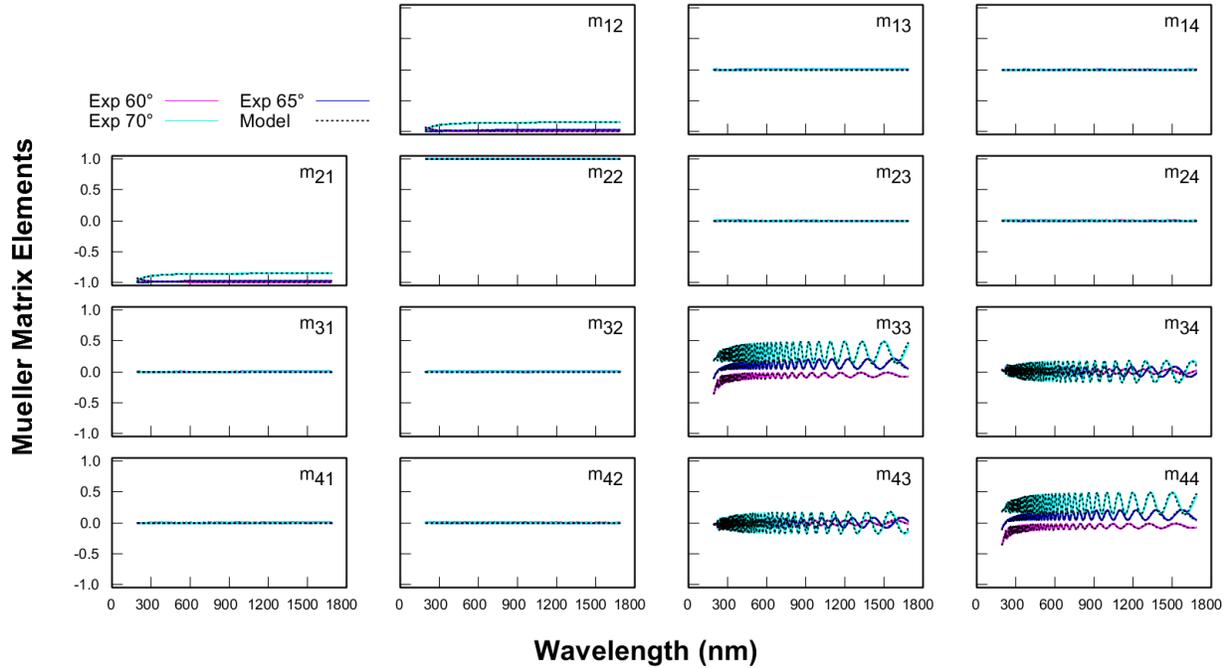

**Figure 6.** Mueller matrix spectroscopic ellipsometry measurement (solid line) and the model-generated data (dotted line) for the 0.273-mm-thick, double-side smooth sample. The data are collected in the reflection mode at 60°, 65°, and 70°.

Ellipsometry is a model-based technique. Determination of the complex and wavelength-dependent refractive index requires the following: i) a presumption of the layered model of the crystal structure of the sample, and ii) the development of the dielectric permittivity function model, which reproduces the measured ellipsometric data. The preliminary tests revealed that the model, in which the sample is represented as a single slab with some roughness at the front interface, is sufficiently versatile to provide enough degrees of freedom to represent the sample geometry. In the case of optically anisotropic material such as SCAM, constructing a valid dielectric permittivity model requires incorporation and simultaneous analysis of the reflection and transmission ellipsometry datasets together with the transmission intensity data (details of the applied procedure can be found in SI). The developed model also incorporates the likely presence of backside reflections[22].

To obtain the dielectric permittivity function, we employ the general oscillator approach, constrained with Kramers-Kronig consistency[35,36]. Depending on the direction of the light propagation with respect to the crystal axes, the developed model is a sum of different types and number of oscillators and poles, each of them being a function of the frequency: $\varepsilon_i = \varepsilon_{\infty,i} + \sum \varepsilon_{poles,i} + \sum \varepsilon_{oscillators,i}$, where $i = x, y, z$. The $\varepsilon_\infty$ term is an offset corresponding to the dielectric constant at infinite frequency. Due to the unusual shape of the absorption edge of SCAM, we have decided to select types of oscillators in the following way. The bandgap region is modeled using the Lorentz and Gaussian functions. The introduction of more complex formulae, such as Tauc-Lorentz or Cody-Lorentz[36], does not further improve the agreement between the measured and model-generated SE data. In the $x$-direction, the model consists of a UV-pole, a Lorentz oscillator, and 3 Gaussian functions (see SI for details of the functions). Two low-energy oscillators (#5 & #6) of 4.206 eV and 0.1 eV are added to take into account the features present in the transmission spectra. Oscillator #5 corresponds to the local minimum reported also in Refs. 32 and 36. With the broadening of 2.5 eV, oscillator #6 gives rise to a small absorption in the NIR region, which was also mentioned in Ref. 32. The UV-pole oscillator represents the high-energy electronic transitions, which do not contribute to absorption in the measured spectral range, but they contribute to a real part of the dielectric function. In the $z$-direction, a simpler composition of the oscillators is enough to represent the experimental data.

We attribute this disproportion to the fact that both ellipsometry and intensity data are less sensitive in that direction. Therefore, we implement the UV and IR poles and the single Gaussian and Lorentz functions. Further limiting the number of oscillators led to a significant increase in the mean squared error (MSE) value. All the parameters are gathered in Table 1 and discussed below in the paragraph devoted to the extinction coefficient. To demonstrate good agreement between the experimental ellipsometry data and the developed model in the entire spectral range, in Figure S5 we show selected (unique) elements from the MM in the reflection and transmission mode, obtained for a 0.273-mm-thick sample (see also the enclosed discussion).

| component of permittivity | # of term | type of term/ oscillator | constant $\varepsilon_{inf}$/ amplitude of the oscillator $A_n$ | broadening of the oscillator $Br_n$ (eV) | centre energy of the oscillator $E_n$ (eV) | centre energy of the oscillator $E_n$ obtained from numerical modelling (eV) |
|---|---|---|---|---|---|---|
| $\varepsilon_{x,y}$ | 1 | $\varepsilon_{inf}$ | $1.5710 \pm 0.0006$ | - | - | - |
| | 2 | $\varepsilon_{UV\ pole}$ | $127.90 \pm 0.02$ | - | $8.649 \pm 0.002$ | - |
| | 3 | $\varepsilon_{Lorentz}$ | $0.0250 \pm 0.0003$ | 0.17 | $7.629 \pm 0.0139$ | 7.66 |
| | 4 | $\varepsilon_{Gaussian}$ | $(217.0 \pm 1.5) \times 10^{-5}$ | $0.422 \pm 0.002$ | $6.586 \pm 0.002$ | 6.63 |
| | 5 | $\varepsilon_{Gaussian}$ | $(2.8 \pm 0.1) \times 10^{-6}$ | 0.5 | 4.206 | - |
| | 6 | $\varepsilon_{Gaussian}$ | $(201 \pm 1) \times 10^{-6}$ | $2.53 \pm 0.01$ | 0.1 | - |
| $\varepsilon_z$ | 1 | $\varepsilon_{inf}$ | $1.6390 \pm 0.0006$ | - | - | - |
| | 2 | $\varepsilon_{UV\ pole}$ | $101.92 \pm 0.02$ | - | $8.716 \pm 0.001$ | - |
| | 3 | $\varepsilon_{Lorentz}$ | $5.574 \pm 0.015$ | $0.1233 \pm 0.0004$ | 6.640 | 7.15 |
| | 4 | $\varepsilon_{Gaussian}$ | $0.0221 \pm 0.0002$ | $0.1140 \pm 0.0005$ | 6.262 | 6.32 |
| | 5 | $\varepsilon_{IR\ pole}$ | $(1574 \pm 6) \times 10^{-6}$ | - | - | - |

**Table 1**. Model parameters of the dielectric permittivity of ScAlMg04 in the sample plane (xy) and out-of-plane direction (z), where $A_n$ is the amplitude, $Br_n$ is the broadening, and $E_n$ is the oscillator's center energy. The error bars provide an estimation of the measurement reproducibility and uniqueness of the specific parameter during the regression analysis and should not be interpreted literally[22]. The parameters given without the error were fixed in the model. The last column shows the oscillator positions extracted from the *ab initio* MBPT.

The developed model is also in line with the measured optical transmission spectra, presented in Figure 7. The overall transmittance is above 80% for wavelengths longer than 400 nm for normal incidence illumination. Similarly, as in previously reported research, the absorption edge appears around 200 nm[32,38]. Above this wavelength, the transmission starts to gradually rise with a change in the slope at 220 nm. Following Ref. 37, we attribute this alteration in the band edge to the electronic transitions, which take place at the localized sites. Moving towards the longer wavelengths, we observe a weak maximum of absorption at the wavelength of 290 nm, represented by oscillator #5 in the model. Although the amplitude of oscillator #5 is almost negligible, it is still possible to extract the oscillator parameters due to the incorporation into our model of the transmission intensity data of a 0.564-mm-thick sample[39]. The peak is not manifested in the Depolarization Index values (DI, see Figure S6), and it disappears after annealing in hydrogen[38]. Moreover, this peak fades out at the higher illumination angles, which means that it contributes to the losses only in the *x*-direction. For the oblique angle of incidence, an additional feature starts to appear in the short-wavelength spectral range (see the inset to Figure 7). It occurs at around 207 nm and is present only when the incident light has a non-zero out-of-plane polarization component. Our model predicts its shape well. We have checked, however, that it does not have a physical meaning and is purely associated with the limited bandwidth of the ellipsometer.

The quality of the proposed model and thus the presented analysis can be validated on the base of the MSE values. Considering the reflection ellipsometry data, together with the transmission intensity and depolarisation data measured for 3 different types of samples, the overall multi-sample MSE equals 11.1. However, if we limit the spectral range to 200-1690 nm, the transmission ellipsometry data are not affected by the depolarization and the MSE drops to 8.9. In the best-fit model, the surface roughness of our SCAM slabs is equal to 2.56 nm for the 0.564-mm-thick sample and 0.79 nm for the 0.273-mm-thick sample.

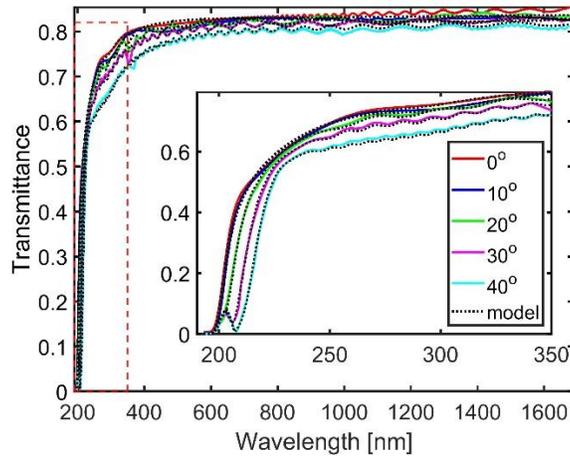

**Figure 7.** Measured transmission intensity spectra of the 0.564-mm-thick SCAM substrate, were obtained for 0° (red solid line), 10° (blue solid line), 20° (green, solid line), 30° (magenta solid line), and 40° (cyan solid line) incidence. The transmission intensity spectra calculated from the model are shown by the black dotted line. The inset shows a close-up of the short-wavelength region of the spectrum.

Finally, we extract from our model the ordinary (in-plane) and extraordinary (out-of-plane) complex refractive index values as a function of the wavelength, as shown in Figure 8. For a comparison with the model based on the B-spline functions, see Figure S7. For the wavelength of 1 μm, the in-plane refractive index of SCAM equals 1.8215, and it is only slightly higher than the ordinary refractive index of sapphire, which is 1.7558. The optical anisotropy of SCAM is present in the entire analyzed spectral range, but it fades out in the short-wavelength range. From the absorption coefficient curve, we also extract the bandgap values determined by a linear extrapolation of the threshold of optical absorption, which equals 6.16 eV and 6.12 eV in the plane of the sample and the out-of-plane direction, respectively. One should note, however, that the precision of determination of these values is lowered by the fact that the SE data barely reach the spectral range above the bandgap.

The above experimental results are worth comparing with the dielectric function, calculated numerically within the RPA LRF formalism described in Section 3. The energy-dependent ordinary and extraordinary complex refractive indices are presented in Figs. 7c and 7d. In addition, extended spectral ranges of the electric permittivity functions are available in Figure S2. The experimental and numerical data qualitatively agree well in the entire spectral range. Although the refractive index values obtained from RPA LRF are slightly smaller than those extracted from the experiment, the mutual correlation between the ordinary and extraordinary components is comparable in both applied methods. The theory predicts not only a fading of the birefringence around 6.32 eV, but also the position of the resonances around the bandgap region (see Table 1). A similar difference between the ordinary and extraordinary refractive indices has been previously reported for sapphire[40], crystalline wurtzite GaN[41], h-BN[42], and $KTiOPO_4$ [43]. It was attributed to a combined contribution of excitonic and inter-band optical transitions, as well as the optically active phonon modes, in which the intensity of both effects was different for the light propagating through the uniaxial crystal in the ordinary and extraordinary direction. A similar explanation should hold for SCAM, in which the crystal structure comprises alternating rock-salt and wurtzite layers stacked along the *c*-axis (see Figure 2).

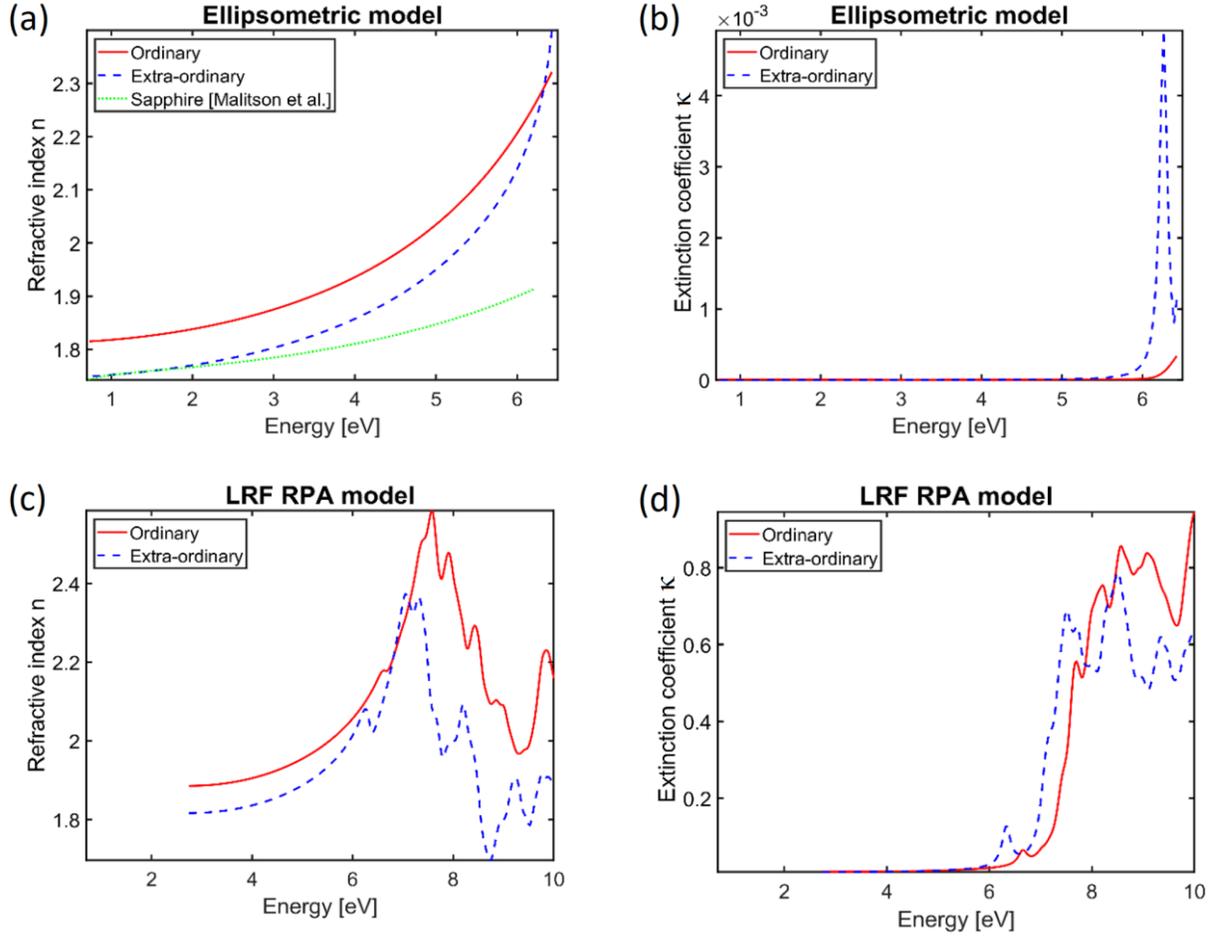

**Figure 8.** (a) and (c) Refractive index, (b) and (d) extinction coefficient extracted from the ellipsometric model (a, b) and from the RPA LRF calculations (c, d). The solid red line corresponds to the in-plane, while the dotted blue line to the out-of-plane directions, respectively. For comparison, the ordinary refractive index of sapphire[40] is also plotted (green line).

As we have demonstrated, SCAM is a highly transparent material in the visible and near-infrared spectral range, with absorption rapidly increasing in the UV. Limiting the analysis of SCAM analysis to the spectral range where losses are negligible, i.e. to the wavelengths longer than 400 nm instead of using a general oscillator model, it is possible to fit ellipsometric data with a simple Cauchy model. The real part of the refractive index can then be expressed by the formulae:

$$n_{in-plane} = 1.811 + \frac{0.00967}{\lambda^2} + \frac{0.00019903}{\lambda^4} \tag{7}$$

$$n_{out-of-plane} = 1.746 + \frac{0.00873}{\lambda^2} + \frac{0.00015422}{\lambda^4} \tag{8}$$

where $n_{in-plane}$ and $n_{out-of-plane}$ are the values for in-plane and out-of-plane light propagation, respectively.

## 4. CONCLUSIONS

Employing various experimental techniques: variable angle spectroscopic ellipsometry, optical transmission, X-ray diffraction, scanning electron microscopy, and Raman spectroscopy, we have provided a comprehensive picture of the optical, electronic and physical properties of single $ScAlMgO_4$ crystals grown by the Czochralski method. In particular, we have established that SCAM is a birefringent material with only a slightly higher refractive index than that of sapphire. We have developed a model based on the ellipsometric data for polarization-resolved dielectric function in a wide UV-VIS-NIR spectral range and have provided Cauchy formulas describing the dispersion of the anisotropic refractive index in the visible spectral range. We support our experimental results with calculations performed in the frame of *ab initio* MBPT. The reported properties are crucial for the design and optimization of optoelectronic and photonic devices incorporating SCAM layers.

Let us note finally, that $ScAlMgO_4$ exhibits an in-plane lattice constant that is very close to lattice constants of transition metal dichalcogenide (TMD) materials. As a result, the lattice mismatch of most commonly studied TMDs to ScAlMgO$_4$ is around an order of magnitude lower than that to hBN that is routinely used as a substrate or encapsulation material for TMD monolayers[44] (see Suppl. Table S2 and references there in [45,46,47,48,49,50,51]). Moreover, ScAlMgO$_4$ might be mechanically exfoliated from the bulk crystal to obtain large, atomically flat flakes like in the case hBN. This indicates that ScAlMgO$_4$ is a perfect substrate for epitaxial growth of TMDs (either in a monolayer or bulk form), encapsulation of exfoliated flakes of TMD monolayers, as well as for studies related to Moire pattern physics[52]. As such, it provides an attractive alternative to hBN.

ASSOCIATED CONTENT

**Supporting Information**

Detailed description of *ab initio* MBPT calculations, ellipsometric measurements, and ellipsometric data processing; Additional figures validating presented analysis of optical properties; Tables with refractive index values and review of the lattice parameters of the SCAM crystals determined by XRD reported in the literature.

**Corresponding Author**

*tomasz.stefaniuk@fuw.edu.pl

ACKNOWLEDGMENT

This work was supported by the Polish National Science Centre within the OPUS projects 2020/37/B/ST8/03446 and 2020/39/B/ST7/03502. The calculations were performed using the Prometheus computer within the PL-Grid supercomputing infrastructure.

# Optical, electronic and structural properties of ScAlMgO$_4$

T. Stefaniuk[a*], J. Suffczyński[a], M. Wierzbowska[b], J. Z. Domagała[c], J. Kisielewski[d], A. Kłos[d], A. Korneluk[a], H. Teisseyre[b,c]

[a] Faculty of Physics, University of Warsaw, 5 Pasteura St., 02-093 Warsaw, Poland

[b] Institute of High Pressure Physics Polish Academy of Sciences, 29 Sokołowska St., 01-142 Warsaw, Poland

[c] Institute of Physics, Polish Academy of Sciences, 32/46 Al. Lotników, 02-668 Warsaw, Poland

[d] Łukasiewicz Research Network - Institute of Microelectronics and Photonics, Wólczyńska 133, 01 919 Warsaw, Poland

* Corresponding author: tomasz.stefaniuk@fuw.edu.pl

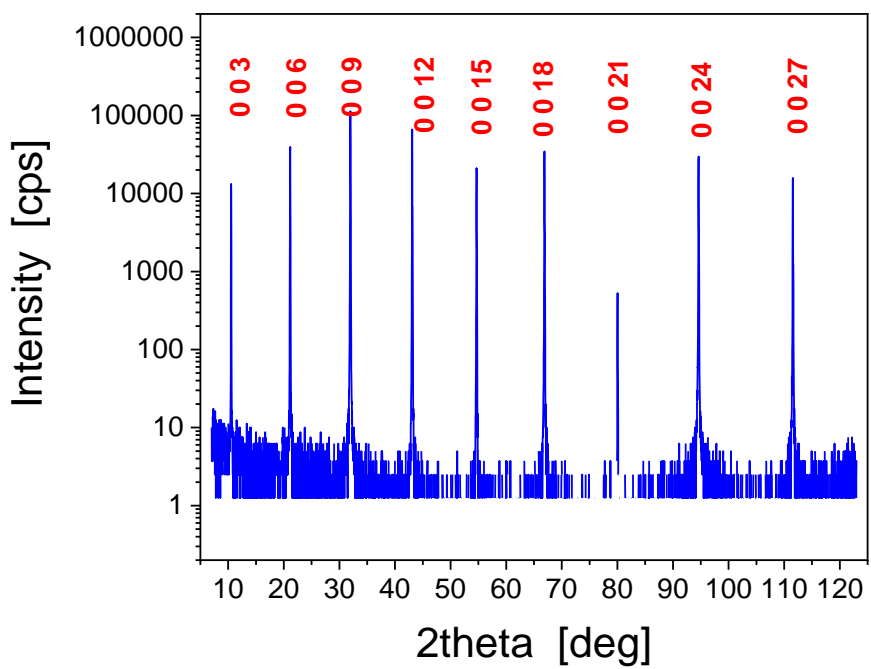

**Figure S1.** 2*theta/omega* scan coupled to the surface [detector preceded by a vertical slit ¼$^0$]; h k l Miller indices are marked in red.

**Additional tables**

| First author | a (Å) | c (Å) | Growth method | Publication year |
|---|---|---|---|---|
| Kimizuka[5] | 3.260 | 25.988 | Powder | 1989 |
| Hellman[1] | 3.2405 | 25.106 | Czochralski | 1996 |
| Ohtomo[3] | 3.249 | 25.195 | Czochralski | 2000 |
| Zhou[9] | 3.2507 | 25.153 | Czochralski | 2009 |
| Mizoguchi[10] | 3.2388 | 25.1056 | Powder; high-temperature solid-state reactions | 2011 |
| **this work** | 3.246 | 25.1352 | Czochralski | 2021 |
| | <3.245±0.005> | <25.14±0.04> | average value and its standard deviation | |

**Table S1**. Review of the lattice parameters of the SCAM crystals determined by XRD reported in the literature with indicated growth method of the sample, publication year, and the name of the first author.

| First author | Material | In-plane lattice constant in Å | Lattice mismatch to | |
|---|---|---|---|---|
| | | | ScAlMgO$_4$ | hBN |
| Agrawal[45] | MoSe$_2$ | 3.283 | -0.0117 | -0.311 |
| Young[46] | MoS$_2$ | 3.148 | 0.0299 | -0.257 |
| Khelil[47] | WeSe$_2$ | 3.290 | -0.0139 | -0.314 |
| Schutte[48] | WS$_2$ | 3.152 | 0.0287 | -0.259 |
| Agarwal[45] | MoTe$_2$ | 3.530 | -0.0878 | -0.410 |
| Schulz[49] | GaN | 3.189 | 0.0173 | -0.274 |
| Reeber[50] | ZnO | 3.250 | -0.0015 | -0.298 |
| average value from Table S1 | ScAlMgO$_4$ | 3.245 | | |
| Lynch[51] | hBN | 2.504 | | |

**Table S2**. In-plane lattice constants of selected semiconductor materials along with the lattice mismatch to ScAlMgO$_4$ and hexagonal boron nitride.

**Detailed equations for linear response function and Bethe-Salpeter equation**

The linear response function is defined as

$$\chi_{\mathbf{GG'}}(\mathbf{q},\omega) = [\delta_{\mathbf{GG''}} - v(\mathbf{q},\mathbf{G''})\chi^0_{\mathbf{GG''}}(\mathbf{q},\omega)]^{-1}\chi^0_{\mathbf{G''G'}}(\mathbf{q},\omega) \tag{S1}$$

while the noninteracting response function $\chi^0_{\mathbf{GG''}}(\mathbf{q},\omega)$ is

$$\chi^0_{\mathbf{GG'}}(\mathbf{q},\omega) = 2\sum_{nn'} \int_{BZ} \frac{d\mathbf{k}}{(2\pi)^3} \rho^*_{n'n\mathbf{k}}(\mathbf{q},\mathbf{G})\rho_{n'n\mathbf{k}}(\mathbf{q},\mathbf{G'})f_{n\mathbf{k}-\mathbf{q}}(1-f_{n'\mathbf{k}}) \\ \times \left[\frac{1}{\omega+\varepsilon_{n\mathbf{k}-\mathbf{q}}-\varepsilon_{n'\mathbf{k}}+i0^+} - \frac{1}{\omega-\varepsilon_{n\mathbf{k}-\mathbf{q}}+\varepsilon_{n'\mathbf{k}}-i0^+}\right] \tag{S2}$$

The above formula uses the DFT eigenvalues $\varepsilon_{n\mathbf{k}}$ as well as the eigenvectors $|n\mathbf{k}\rangle$, the latter being embedded in Eq. (4) via the screening-matrix elements $\rho_{nn'\mathbf{k}}$ defined as

$$\rho_{nn'\mathbf{k}}(\mathbf{q},\mathbf{G}) = \langle n\mathbf{k}|e^{i(\mathbf{q}+\mathbf{G})\mathbf{r}}|n'\mathbf{k}-\mathbf{q}\rangle \tag{S3}$$

The wave function is expanded over the reciprocal lattice vectors

$$|n\mathbf{k}\rangle = \Phi_{n\mathbf{k}}(\mathbf{r}) = \frac{1}{V^{1/2}}\sum_{\mathbf{G}} e^{i(\mathbf{k}+\mathbf{G})\mathbf{r}}c_{n\mathbf{k}}(\mathbf{G}) \tag{S4}$$

and $f_{n\mathbf{k}}$ are Fermi-Dirac distribution occupation numbers of the band $n$ at the k-point in BZ.

The Bethe-Salpeter equation takes into account the effect of the electron-hole (e-h) interactions via the noninteracting Green's function $L^0$, which enters

$$\lim_{\mathbf{q}\to 0}\chi^0_{\mathbf{GG'}}(\mathbf{q},\omega) = -\sum_{nn'\mathbf{k}} \lim_{\mathbf{q}\to 0}[\rho^*_{n'n\mathbf{k}}(\mathbf{q},\mathbf{G})\rho_{nn'\mathbf{k}}(\mathbf{q},\mathbf{G'})]L^0_{nn'\mathbf{k}}(\omega) \tag{S5}$$

Thus, the interacting polarization takes the form

$$\lim_{\mathbf{q}\to 0}\tilde{\chi}_{\mathbf{GG'}}(\mathbf{q},\omega) = -\sum_{nn'\mathbf{k}}\sum_{mm'\mathbf{k'}} \lim_{\mathbf{q}\to 0}[\rho^*_{n'n\mathbf{k}}(\mathbf{q},\mathbf{G})\rho_{m'm\mathbf{k'}}(\mathbf{q},\mathbf{G'})]\tilde{L}_{nn'\mathbf{k},mm'\mathbf{k'}}(\omega) \tag{S6}$$

The Bethe-Salpeter equation (BSE) is an equation for $L_{nn'k,mm'k'}$ and reads

$$\tilde{L}_{nn'\mathbf{k},mm'\mathbf{k'}}(\omega) = L^0_{nn'\mathbf{k}}(\omega)[\delta_{nm}\delta_{n'm'}\delta_{\mathbf{kk'}} + i\sum_{ss'k_1} \Xi_{nn'k,ss'k_1}\tilde{L}_{ss'\mathbf{k_1},mm'\mathbf{k'}}(\omega)] \tag{S7}$$

with the self energy $\Xi_{nn'k,ss'k_1}$ defined as

$$\Xi_{nn'k,ss'k_1} = W_{nn'k,ss'k_1} - 2\tilde{V}_{nn'k,ss'k_1} \tag{S8}$$

while W and V are integrals, which employ the DFT Bloch functions via the screening-matrix elements $\rho$, given in Eqs. (3) and (4), as follows

$$W_{nn'k,ss'k_1} = \frac{1}{\Omega N_\mathbf{q}}\sum_{\mathbf{GG'}} \rho_{ns}(\mathbf{k},\mathbf{q}=\mathbf{k}-\mathbf{k_1},\mathbf{G})\rho^*_{n's'}(\mathbf{k_1},\mathbf{q}=\mathbf{k}-\mathbf{k_1},\mathbf{G'})\varepsilon^{-1}_{\mathbf{GG'}}v(\mathbf{q},\mathbf{G'}) \tag{S9}$$

$$\tilde{V}_{nn'k,ss'k_1} = \frac{1}{\Omega N_\mathbf{q}}\sum_{\mathbf{G}\neq 0} \rho_{nn'}(\mathbf{k},\mathbf{q}=0,\mathbf{G})\rho^*_{ss'}(\mathbf{k_1},\mathbf{q}=0,\mathbf{G})v(\mathbf{G}) \tag{S10}$$

where $N_\mathbf{q}$ is the number of points in the BZ sampling, and $v(\mathbf{q} + \mathbf{G})$ is the bare Coulomb interaction in the reciprocal space. For the system with the energy gap and at zero temperature, the noninteracting Green's function $L^0$ is

$$\mathcal{R}[L^0_{nn'\mathbf{k}}(\omega)] = i\frac{f_{n'\mathbf{k}} - f_{n\mathbf{k}}}{\omega - \varepsilon_{n'\mathbf{k}} + \varepsilon_{n\mathbf{k}}} \quad (S11)$$

and the BSE simplifies to the eigenvalue problem with the Hamiltonian

$$H_{nn'\mathbf{k},mm'\mathbf{k}'} = (\varepsilon_{n'\mathbf{k}} - \varepsilon_{n\mathbf{k}})\delta_{nm}\delta_{n'm'}\delta_{\mathbf{k}\mathbf{k}'} + (f_{n'\mathbf{k}} - f_{n\mathbf{k}})[2\tilde{V}_{nn'\mathbf{k},mm'\mathbf{k}'} - W_{nn'\mathbf{k},mm'\mathbf{k}'}] \quad (S12)$$

with the DFT eigenvalues $\varepsilon_{n\mathbf{k}}$. In the above equation, the unoccupied states might be shifted up by the value corresponding to the correction of the DFT bandgap obtained with the GW method.

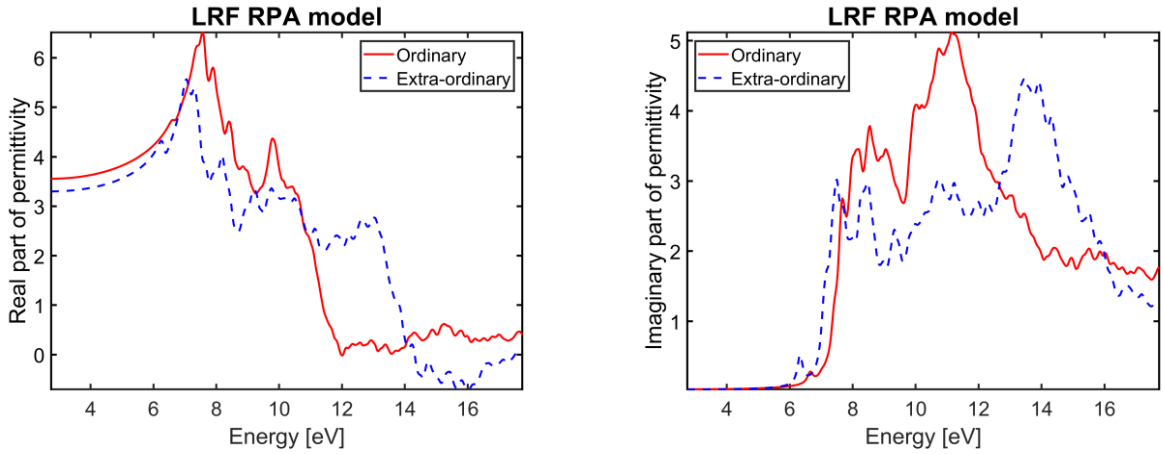

**Figure S2.** (a) The real part of the macroscopic dielectric function calculated with the RPA linear response function for 2 polarizations of the electric-field vector: perpendicular to it (ordinary) and parallel to the crystal $c$-axis (extraordinary). (b) The imaginary part of the macroscopic dielectric function.

**Ellipsometric data analysis procedure**

The model for the dielectric permittivity function is constructed in multiple steps. First, we analyze the reflection MM data obtained from the backside roughened sample #3 and create an initial model. This data set carries information mainly about the ordinary refractive index of the material and the sample roughness. But it contains almost no information about the sample thickness, losses, or extraordinary refractive index. In the second step, we incorporate the reflection MM data obtained from the double-side smooth samples #1 and #2. Simultaneous analysis of the samples with different thicknesses helps to remove any ambiguity from the model. Due to the oscillations of the values of the diagonal elements of the MM, these datasets are sensitive to the difference between ordinary and extraordinary rays in the refractive index values. The smooth back surface of the samples gives rise to the backside reflections. Because the backside reflections might lead to a depolarization of the detected signal and affect the MM components, the presence of the back-reflected light was also included in the model as well as the depolarization data. This aspect will be further discussed. In the third step, we include the transmission intensity data (obtained from samples #1 and #2) in the model. This step allows for precise estimation of the imaginary part of the refractive index and helps set the samples' thickness. Finally, we add the transmission MM data into the constructed model (from samples #1 and #2). This dataset, obtained at high illumination angles, contains information about the absolute refractive index values and retardance. All the above-described datasets are combined within the so-called multi-sample model[22]. At the same time, the mean square error (MSE) values are calculated as the difference between the measured data and the obtained phenomenological curves, based on the knowledge of the MM dataset in the entire spectral range (under the assumption that all data points in the spectrum contain meaningful information).

**Details of the general oscillator model**
Depending on the direction of the light propagation with respect to the crystal axes, the developed model is the sum of a different number of oscillators and their types and poles. The particular terms included in the developed model for the permittivity are expressed by the following equations:

- UV pole:

$$\varepsilon(E) = \frac{A_n}{E_n^2 - E^2}, \tag{S13}$$

- IR pole:

$$\varepsilon(E) = \frac{-A_n}{E^2}, \tag{S14}$$

- Lorentz

$$\varepsilon(E) = \frac{A_n Br_n E_n}{E_n^2 - E^2 - iBr_n E}, \tag{S15}$$

- Gaussian

$$\varepsilon(E) = A_n \left\{ \left[ \Gamma\left(\frac{E - E_n}{\sigma_n}\right) + \Gamma\left(\frac{E + E_n}{\sigma_n}\right) \right] + i \left( exp\left[ -\left(\frac{E - E_n}{\sigma_n}\right)^2 \right] - exp\left[ -\left(\frac{E + E_n}{\sigma_n}\right)^2 \right] \right) \right\}, \tag{S16a}$$

$$\sigma_n = \frac{Br_n}{2\sqrt{ln(2)}}. \tag{S16b}$$

where $n$ is the oscillator number, $A_n$ is the amplitude, $Br_n$ is the broadening, and $E_n$ is the oscillator's central energy. The function $\Gamma$ is introduced as a flexible parameter to achieve a convergent series of results that reproduce the line shapes of the real part of the permittivity such that they are consistent with the Kramers-Kronig relations.

The non-linear Levenberg-Marquardt algorithm was used to retrieve the best-fitting parameters. In the most general case, when all oscillators and pole parameters are free, we observe strong correlations in the x-direction between the amplitude and broadening of the #3 oscillator and the amplitude and position of the #6 oscillator (see Table 1 in the manuscript). This can be quite easily understood because both resonances are located outside the measurement range of the ellipsometer. Thus, we decided to fix the value of the width of the first one and the central energy of the latter one.

To demonstrate good agreement between the experimental ellipsometry data and the developed model, we show selected elements from the MM in the reflection and transmission mode obtained for the 0.273-mm-thick sample (see Figure S5). The sample is vertically mounted to the ellipsometer stage using a vacuum-based system, and a possible tilt of the sample is corrected in reflection at a 65° angle. For the reflection, the model correctly describes the wavelength dependence of the $m_{21}$ element, which corresponds to the linear attenuation between the horizontal and vertical linear polarization, as well as the $m_{33}$ and $m_{34}$ elements connected to the retardance of the ordinary refractive index value. The model reproduces the oscillations' amplitude, modulation, and period (present in $m_{33}$ and $m_{34}$ elements). It can also predict the shape of the feature at the wavelength of 234 nm and the disappearance of the oscillations for the wavelengths shorter than 224 nm. A very good match is also obtained for the data gathered in the transmission mode. The only deviation exists for the $m_{21}$ term, and only for a very narrow spectral range of 193-200 nm and high illumination angles. This disagreement, however, can be explained on the basis of the depolarization index values (DI, see Figure S6), as discussed below.

In the reflection mode, the fluctuations of the DI values reflect the oscillations present in the MM elements, because the depolarization also originates from the backside reflections[39]. Close to unity DI values, and the fact that the model explains well the shape of the DI, indicates that the depolarization does not deteriorate the ellipsometric data, obtained in reflection mode. In contrast, in the transmission mode in UV, we observe a large increase of the depolarization (i.e. a decrease of the DI values) for the wavelengths shorter than 230 nm. The origin of this phenomenon is not clear. An in-depth analysis of the model shows that the depolarization in this spectral range is not caused by thickness inhomogeneity, surface scattering, incidence angle variation, or backside reflections. However, because the created model predicts the trends in the DI curve, we believe that the increase of the DI values is linked with the vicinity of the bandgap resonances.

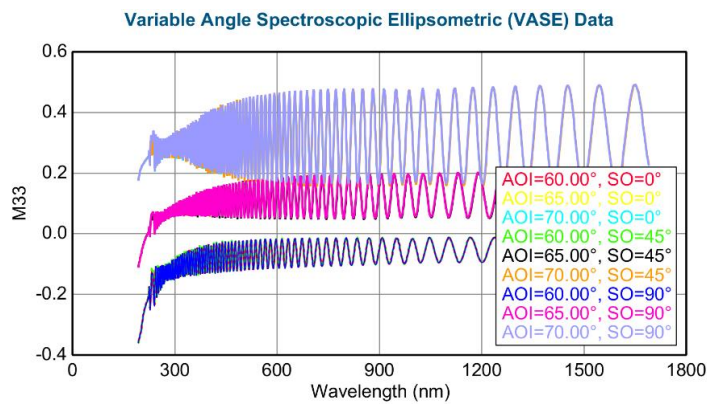

**Figure S3.** The M33 element of Mueller matrix spectroscopic ellipsometry measurement obtained in the reflection mode for the 0.564-mm-thick SCAM sample for different angles of incidence (AOI) and different sample orientations (SO). SO angle refers to the rotation around the *z*-direction (i.e. out-of-plane direction).

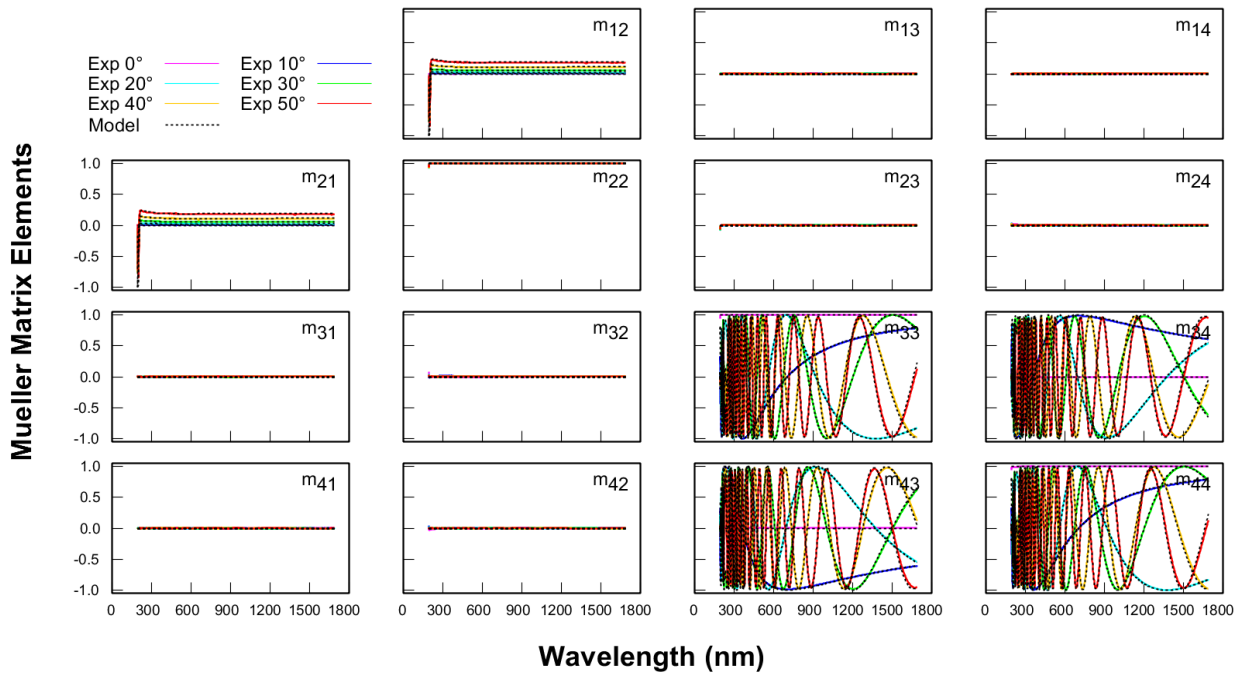

**Figure S4.** Mueller matrix spectroscopic ellipsometry measurement (solid line) and model-generated (dotted line) data for the 0.273-mm-thick double-side smooth sample, in the transmission mode for the angle of incidence in a range from 0° to 50° with a step of 10°.

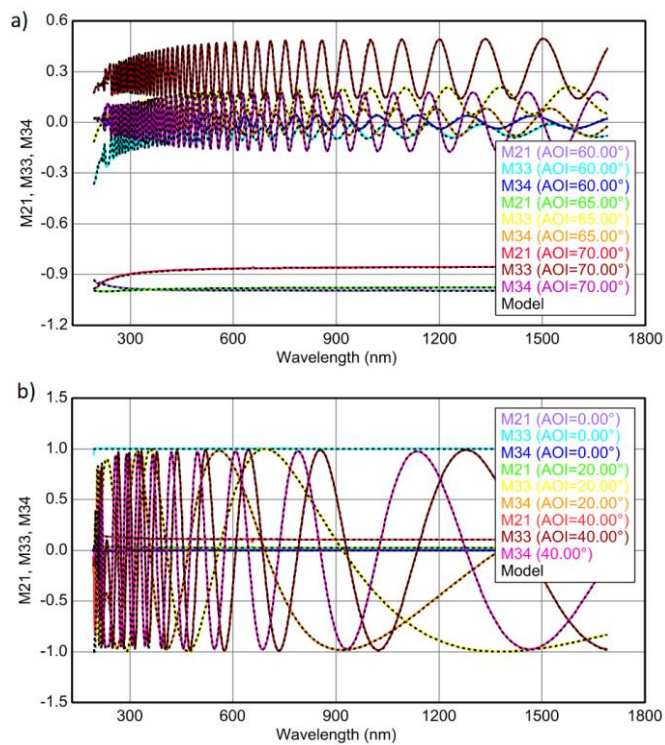

**Figure S5.** Selected (unique) elements from the Mueller matrix were measured in (a) the reflection mode and (b) the transmission mode for different angles of incidence (AOI). The data correspond to a sample with the thickness of 0.273 mm.

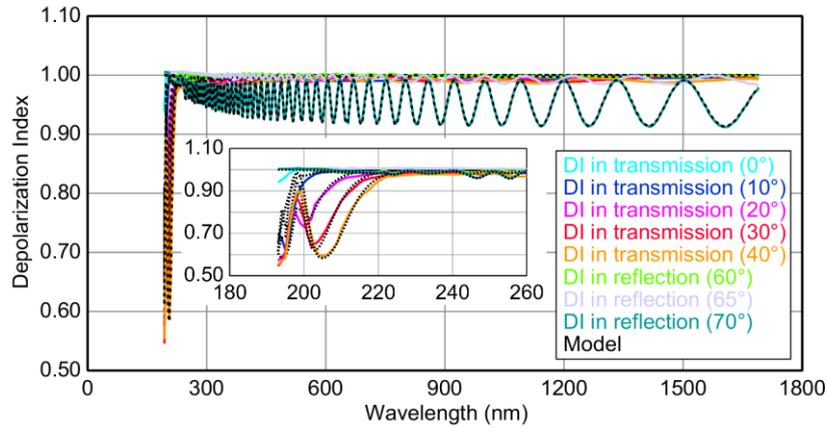

**Figure S6.** Depolarization Index (DI) calculated from the Mueller matrix in reflection and transmission mode. The inset presents a close-up of the UV region. The data correspond to a sample with the thickness of 0.273 mm.

**B-spline model**

To determine how the construction of the model influences the values of the optical parameters retrieved for the material at hand, we compared the refractive index values estimated on the basis of the proposed general oscillator model with the ones coming from the fitting of a series of the B-spline functions to the experimental datasets. In the latter approach, by summing the B-spline basis functions together with the weighted amplitudes of each node, one can obtain the resulting permittivity curve fulfilling the condition of Kramers-Kronig consistency. Thanks to the possibility of tuning the approximate spacing of the nodes, the method provides greater flexibility than the general oscillator model, regarding the shape of the refractive index functions. The ordinary and extraordinary refractive indices of SCAM are described with the B-spline consisting of 89 nodes each. A node resolution of 0.1 eV is used in the transparent spectral regions, while a finer resolution of 0.05 eV is used in the UV range (193-400 nm). A comparison of the results obtained from the general oscillator model and the B-spline functions is presented in Figure S7. We achieve excellent agreement, which proves that the developed general oscillator model can indeed reproduce the exact shape of the refractive index functions.

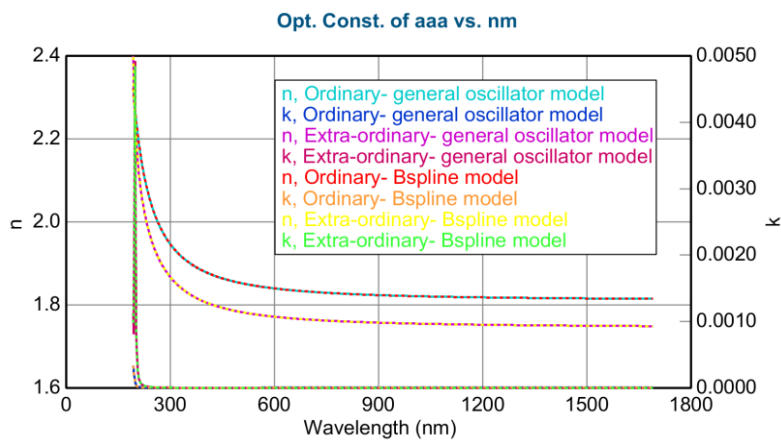

**Figure S7.** Comparison of real and imaginary parts of the refractive index values estimated using either a general oscillator or the B-spline model.

**Values of refractive index of $ScAlMgO_4$**

| | Refractive index of $ScAlMgO_4$ | | | |
|---|---|---|---|---|
| Wavelength (nm) | The real part of ordinary refractive index | The imaginary part of the ordinary refractive index | The real part of extraordinary refractive index | The imaginary part of the extraordinary refractive index |
| 193 | 2.32146 | 0.000338 | 2.399775 | 0.001125 |
| 194 | 2.311172 | 0.000285 | 2.353716 | 0.000808 |
| 195 | 2.301235 | 0.000235 | 2.318696 | 0.001217 |
| 196 | 2.291634 | 0.000188 | 2.290442 | 0.002093 |
| 197 | 2.282353 | 0.000148 | 2.267291 | 0.003651 |
| 198 | 2.273377 | 0.000114 | 2.249111 | 0.004919 |
| 199 | 2.264694 | 0.000087 | 2.233522 | 0.003874 |
| 200 | 2.256292 | 0.000066 | 2.218109 | 0.002315 |
| 201 | 2.248158 | 0.00005 | 2.203532 | 0.001391 |
| 202 | 2.240281 | 0.000039 | 2.190122 | 0.000898 |
| 203 | 2.232649 | 0.000031 | 2.177796 | 0.00062 |
| 204 | 2.225251 | 0.000025 | 2.166404 | 0.000453 |
| 205 | 2.218078 | 0.000021 | 2.15581 | 0.000345 |
| 206 | 2.211119 | 0.000018 | 2.145903 | 0.000271 |
| 207 | 2.204365 | 0.000017 | 2.136596 | 0.00022 |
| 208 | 2.197807 | 0.000015 | 2.127818 | 0.000182 |
| 209 | 2.191436 | 0.000014 | 2.119508 | 0.000153 |
| 210 | 2.185246 | 0.000014 | 2.111618 | 0.000131 |
| 211 | 2.179227 | 0.000013 | 2.104107 | 0.000113 |
| 212 | 2.173374 | 0.000013 | 2.09694 | 0.000099 |
| 213 | 2.167679 | 0.000012 | 2.090088 | 0.000087 |
| 214 | 2.162137 | 0.000012 | 2.083523 | 0.000078 |
| 215 | 2.156741 | 0.000012 | 2.077224 | 0.00007 |
| 216 | 2.151486 | 0.000011 | 2.071171 | 0.000063 |
| 217 | 2.146366 | 0.000011 | 2.065346 | 0.000057 |
| 218 | 2.141376 | 0.000011 | 2.059733 | 0.000053 |
| 219 | 2.136513 | 0.00001 | 2.054319 | 0.000048 |
| 220 | 2.13177 | 0.00001 | 2.049091 | 0.000044 |
| 221 | 2.127143 | 0.00001 | 2.044038 | 0.000041 |
| 222 | 2.122629 | 0.00001 | 2.03915 | 0.000038 |
| 223 | 2.118224 | 0.000009 | 2.034417 | 0.000036 |
| 224 | 2.113923 | 0.000009 | 2.02983 | 0.000033 |
| 225 | 2.109724 | 0.000009 | 2.025383 | 0.000031 |
| 226 | 2.105622 | 0.000009 | 2.021068 | 0.000029 |
| 227 | 2.101614 | 0.000009 | 2.016878 | 0.000028 |
| 228 | 2.097698 | 0.000008 | 2.012807 | 0.000026 |
| 229 | 2.09387 | 0.000008 | 2.00885 | 0.000025 |
| 230 | 2.090128 | 0.000008 | 2.005001 | 0.000023 |
| 231 | 2.086468 | 0.000008 | 2.001256 | 0.000022 |
| 232 | 2.082888 | 0.000008 | 1.997609 | 0.000021 |
| 233 | 2.079386 | 0.000008 | 1.994058 | 0.00002 |
| 234 | 2.075959 | 0.000008 | 1.990597 | 0.000019 |
| 235 | 2.072605 | 0.000007 | 1.987223 | 0.000018 |
| 236 | 2.069321 | 0.000007 | 1.983933 | 0.000018 |
| 237 | 2.066106 | 0.000007 | 1.980723 | 0.000017 |
| 238 | 2.062957 | 0.000007 | 1.97759 | 0.000016 |
| 239 | 2.059873 | 0.000007 | 1.974532 | 0.000016 |
| 240 | 2.056851 | 0.000007 | 1.971545 | 0.000015 |
| 241 | 2.05389 | 0.000007 | 1.968626 | 0.000014 |
| 242 | 2.050988 | 0.000007 | 1.965775 | 0.000014 |
| 243 | 2.048143 | 0.000006 | 1.962987 | 0.000013 |
| 244 | 2.045354 | 0.000006 | 1.960261 | 0.000013 |
| 245 | 2.042619 | 0.000006 | 1.957594 | 0.000013 |
| 246 | 2.039936 | 0.000006 | 1.954986 | 0.000012 |
| 247 | 2.037305 | 0.000006 | 1.952433 | 0.000012 |
| 248 | 2.034724 | 0.000006 | 1.949934 | 0.000011 |
| 249 | 2.032191 | 0.000006 | 1.947488 | 0.000011 |
| 250 | 2.029705 | 0.000006 | 1.945092 | 0.000011 |
| 251 | 2.027265 | 0.000006 | 1.942744 | 0.00001 |
| 252 | 2.02487 | 0.000006 | 1.940445 | 0.00001 |
| 253 | 2.022518 | 0.000006 | 1.938191 | 0.00001 |
| 254 | 2.020209 | 0.000005 | 1.935982 | 0.00001 |
| 255 | 2.017941 | 0.000005 | 1.933817 | 0.000009 |
| 256 | 2.015713 | 0.000005 | 1.931693 | 0.000009 |
| 257 | 2.013524 | 0.000005 | 1.92961 | 0.000009 |
| 258 | 2.011374 | 0.000005 | 1.927567 | 0.000009 |
| 259 | 2.009261 | 0.000005 | 1.925563 | 0.000008 |
| 260 | 2.007184 | 0.000005 | 1.923595 | 0.000008 |
| 261 | 2.005143 | 0.000005 | 1.921665 | 0.000008 |
| 262 | 2.003137 | 0.000005 | 1.919769 | 0.000008 |
| 263 | 2.001164 | 0.000005 | 1.917908 | 0.000008 |
| 264 | 1.999224 | 0.000005 | 1.916081 | 0.000008 |
| 265 | 1.997317 | 0.000005 | 1.914286 | 0.000007 |
| 266 | 1.995441 | 0.000005 | 1.912523 | 0.000007 |
| 267 | 1.993595 | 0.000005 | 1.910791 | 0.000007 |
| 268 | 1.99178 | 0.000005 | 1.909089 | 0.000007 |
| 269 | 1.989994 | 0.000005 | 1.907416 | 0.000007 |
| 270 | 1.988236 | 0.000005 | 1.905772 | 0.000007 |
| 271 | 1.986507 | 0.000005 | 1.904156 | 0.000006 |
| 272 | 1.984805 | 0.000005 | 1.902567 | 0.000006 |
| 273 | 1.98313 | 0.000005 | 1.901005 | 0.000006 |
| 274 | 1.98148 | 0.000005 | 1.899468 | 0.000006 |
| 275 | 1.979857 | 0.000005 | 1.897957 | 0.000006 |
| 276 | 1.978258 | 0.000005 | 1.896471 | 0.000006 |

| 277 | 1.976684 | 0.000004 | 1.895008 | 0.000006 |
| 278 | 1.975134 | 0.000004 | 1.893569 | 0.000006 |
| 279 | 1.973607 | 0.000004 | 1.892154 | 0.000006 |
| 280 | 1.972104 | 0.000004 | 1.89076 | 0.000005 |
| 281 | 1.970622 | 0.000004 | 1.889389 | 0.000005 |
| 282 | 1.969163 | 0.000004 | 1.888038 | 0.000005 |
| 283 | 1.967725 | 0.000004 | 1.886709 | 0.000005 |
| 284 | 1.966308 | 0.000004 | 1.885401 | 0.000005 |
| 285 | 1.964912 | 0.000004 | 1.884112 | 0.000005 |
| 286 | 1.963536 | 0.000004 | 1.882843 | 0.000005 |
| 287 | 1.96218 | 0.000004 | 1.881593 | 0.000005 |
| 288 | 1.960843 | 0.000004 | 1.880362 | 0.000005 |
| 289 | 1.959526 | 0.000004 | 1.879149 | 0.000005 |
| 290 | 1.958227 | 0.000004 | 1.877954 | 0.000005 |
| 291 | 1.956946 | 0.000004 | 1.876776 | 0.000005 |
| 292 | 1.955683 | 0.000004 | 1.875616 | 0.000005 |
| 293 | 1.954437 | 0.000004 | 1.874473 | 0.000004 |
| 294 | 1.953209 | 0.000004 | 1.873346 | 0.000004 |
| 295 | 1.951998 | 0.000004 | 1.872235 | 0.000004 |
| 296 | 1.950803 | 0.000004 | 1.87114 | 0.000004 |
| 297 | 1.949624 | 0.000004 | 1.87006 | 0.000004 |
| 298 | 1.948462 | 0.000004 | 1.868996 | 0.000004 |
| 299 | 1.947315 | 0.000004 | 1.867947 | 0.000004 |
| 300 | 1.946183 | 0.000004 | 1.866912 | 0.000004 |
| 301 | 1.945066 | 0.000004 | 1.865891 | 0.000004 |
| 302 | 1.943964 | 0.000004 | 1.864884 | 0.000004 |
| 303 | 1.942877 | 0.000004 | 1.863892 | 0.000004 |
| 304 | 1.941803 | 0.000004 | 1.862912 | 0.000004 |
| 305 | 1.940744 | 0.000004 | 1.861946 | 0.000004 |
| 306 | 1.939698 | 0.000004 | 1.860993 | 0.000004 |
| 307 | 1.938666 | 0.000004 | 1.860052 | 0.000004 |
| 308 | 1.937647 | 0.000004 | 1.859124 | 0.000004 |
| 309 | 1.936641 | 0.000004 | 1.858208 | 0.000004 |
| 310 | 1.935648 | 0.000004 | 1.857304 | 0.000004 |
| 311 | 1.934667 | 0.000004 | 1.856412 | 0.000004 |
| 312 | 1.933699 | 0.000004 | 1.855532 | 0.000004 |
| 313 | 1.932742 | 0.000003 | 1.854662 | 0.000003 |
| 314 | 1.931798 | 0.000003 | 1.853804 | 0.000003 |
| 315 | 1.930865 | 0.000003 | 1.852957 | 0.000003 |
| 316 | 1.929943 | 0.000003 | 1.852121 | 0.000003 |
| 317 | 1.929033 | 0.000003 | 1.851295 | 0.000003 |
| 318 | 1.928134 | 0.000003 | 1.850479 | 0.000003 |
| 319 | 1.927246 | 0.000003 | 1.849674 | 0.000003 |
| 320 | 1.926369 | 0.000003 | 1.848879 | 0.000003 |
| 321 | 1.925502 | 0.000003 | 1.848093 | 0.000003 |
| 322 | 1.924645 | 0.000003 | 1.847317 | 0.000003 |
| 323 | 1.923799 | 0.000003 | 1.846551 | 0.000003 |
| 324 | 1.922962 | 0.000003 | 1.845794 | 0.000003 |
| 325 | 1.922135 | 0.000003 | 1.845046 | 0.000003 |
| 326 | 1.921319 | 0.000003 | 1.844307 | 0.000003 |
| 327 | 1.920511 | 0.000003 | 1.843577 | 0.000003 |
| 328 | 1.919713 | 0.000003 | 1.842856 | 0.000003 |
| 329 | 1.918924 | 0.000003 | 1.842143 | 0.000003 |
| 330 | 1.918145 | 0.000003 | 1.841439 | 0.000003 |
| 331 | 1.917374 | 0.000003 | 1.840742 | 0.000003 |
| 332 | 1.916612 | 0.000003 | 1.840054 | 0.000003 |
| 333 | 1.915858 | 0.000003 | 1.839374 | 0.000003 |
| 334 | 1.915113 | 0.000003 | 1.838702 | 0.000003 |
| 335 | 1.914377 | 0.000003 | 1.838038 | 0.000003 |
| 336 | 1.913648 | 0.000003 | 1.837381 | 0.000003 |
| 337 | 1.912928 | 0.000003 | 1.836732 | 0.000003 |
| 338 | 1.912216 | 0.000003 | 1.83609 | 0.000003 |
| 339 | 1.911512 | 0.000003 | 1.835456 | 0.000003 |
| 340 | 1.910815 | 0.000003 | 1.834828 | 0.000003 |
| 341 | 1.910126 | 0.000003 | 1.834208 | 0.000003 |
| 342 | 1.909445 | 0.000003 | 1.833594 | 0.000003 |
| 343 | 1.908771 | 0.000003 | 1.832987 | 0.000003 |
| 344 | 1.908104 | 0.000003 | 1.832387 | 0.000003 |
| 345 | 1.907444 | 0.000003 | 1.831794 | 0.000003 |
| 346 | 1.906792 | 0.000003 | 1.831207 | 0.000003 |

| 347 | 1.906146 | 0.000003 | 1.830627 | 0.000003 |
| --- | --- | --- | --- | --- |
| 348 | 1.905507 | 0.000003 | 1.830052 | 0.000002 |
| 349 | 1.904876 | 0.000003 | 1.829484 | 0.000002 |
| 350 | 1.90425 | 0.000003 | 1.828923 | 0.000002 |
| 351 | 1.903631 | 0.000003 | 1.828367 | 0.000002 |
| 352 | 1.903019 | 0.000003 | 1.827817 | 0.000002 |
| 353 | 1.902413 | 0.000003 | 1.827273 | 0.000002 |
| 354 | 1.901814 | 0.000003 | 1.826734 | 0.000002 |
| 355 | 1.90122 | 0.000002 | 1.826202 | 0.000002 |
| 356 | 1.900633 | 0.000002 | 1.825675 | 0.000002 |
| 357 | 1.900052 | 0.000002 | 1.825153 | 0.000002 |
| 358 | 1.899476 | 0.000002 | 1.824637 | 0.000002 |
| 359 | 1.898907 | 0.000002 | 1.824126 | 0.000002 |
| 360 | 1.898343 | 0.000002 | 1.823621 | 0.000002 |
| 361 | 1.897785 | 0.000002 | 1.82312 | 0.000002 |
| 362 | 1.897232 | 0.000002 | 1.822625 | 0.000002 |
| 363 | 1.896685 | 0.000002 | 1.822135 | 0.000002 |
| 364 | 1.896144 | 0.000002 | 1.82165 | 0.000002 |
| 365 | 1.895608 | 0.000002 | 1.82117 | 0.000002 |
| 366 | 1.895077 | 0.000002 | 1.820694 | 0.000002 |
| 367 | 1.894551 | 0.000002 | 1.820223 | 0.000002 |
| 368 | 1.894031 | 0.000002 | 1.819757 | 0.000002 |
| 369 | 1.893516 | 0.000002 | 1.819296 | 0.000002 |
| 370 | 1.893005 | 0.000002 | 1.818839 | 0.000002 |
| 371 | sty.25 | 0.000002 | 1.818387 | 0.000002 |
| 372 | 1.891999 | 0.000002 | 1.817939 | 0.000002 |
| 373 | 1.891504 | 0.000002 | 1.817496 | 0.000002 |
| 374 | 1.891013 | 0.000002 | 1.817057 | 0.000002 |
| 375 | 1.890527 | 0.000002 | 1.816622 | 0.000002 |
| 376 | 1.890045 | 0.000002 | 1.816191 | 0.000002 |
| 377 | 1.889568 | 0.000002 | 1.815765 | 0.000002 |
| 378 | 1.889096 | 0.000002 | 1.815343 | 0.000002 |
| 379 | 1.888628 | 0.000002 | 1.814924 | 0.000002 |
| 380 | 1.888164 | 0.000002 | 1.81451 | 0.000002 |
| 381 | 1.887704 | 0.000002 | sty.41 | 0.000002 |
| 382 | 1.887249 | 0.000002 | 1.813693 | 0.000002 |
| 383 | 1.886799 | 0.000002 | 1.81329 | 0.000002 |
| 384 | 1.886352 | 0.000002 | 1.812892 | 0.000002 |
| 385 | 1.885909 | 0.000002 | 1.812496 | 0.000002 |
| 386 | 1.885471 | 0.000002 | 1.812105 | 0.000002 |
| 387 | 1.885037 | 0.000002 | 1.811717 | 0.000002 |
| 388 | 1.884606 | 0.000002 | 1.811333 | 0.000002 |
| 389 | 1.884179 | 0.000002 | 1.810952 | 0.000002 |
| 390 | 1.883757 | 0.000002 | 1.810575 | 0.000002 |
| 391 | 1.883338 | 0.000002 | 1.810201 | 0.000002 |
| 392 | 1.882923 | 0.000002 | 1.809831 | 0.000002 |
| 393 | 1.882511 | 0.000002 | 1.809464 | 0.000002 |
| 394 | 1.882104 | 0.000002 | 1.809101 | 0.000002 |
| 395 | sty.17 | 0.000002 | 1.80874 | 0.000002 |
| 396 | 1.881299 | 0.000002 | 1.808383 | 0.000002 |
| 397 | 1.880902 | 0.000002 | 1.808029 | 0.000002 |
| 398 | 1.880509 | 0.000002 | 1.807678 | 0.000002 |
| 399 | 1.880119 | 0.000002 | 1.807331 | 0.000002 |
| 400 | 1.879732 | 0.000002 | 1.806986 | 0.000002 |
| 401 | 1.879349 | 0.000002 | 1.806645 | 0.000002 |
| 402 | 1.878969 | 0.000002 | 1.806306 | 0.000002 |
| 403 | 1.878592 | 0.000002 | 1.805971 | 0.000002 |
| 404 | 1.878219 | 0.000002 | 1.805638 | 0.000002 |
| 405 | 1.877849 | 0.000002 | 1.805309 | 0.000002 |
| 406 | 1.877482 | 0.000002 | 1.804982 | 0.000002 |
| 407 | 1.877118 | 0.000002 | 1.804658 | 0.000002 |
| 408 | 1.876758 | 0.000002 | 1.804337 | 0.000002 |
| 409 | sty.64 | 0.000002 | 1.804018 | 0.000002 |
| 410 | 1.876046 | 0.000002 | 1.803703 | 0.000002 |
| 411 | 1.875694 | 0.000002 | 1.80339 | 0.000002 |
| 412 | 1.875345 | 0.000002 | 1.80308 | 0.000002 |
| 413 | 1.875 | 0.000002 | 1.802772 | 0.000002 |
| 414 | 1.874657 | 0.000002 | 1.802467 | 0.000002 |
| 415 | 1.874317 | 0.000002 | 1.802165 | 0.000002 |
| 416 | 1.87398 | 0.000002 | 1.801865 | 0.000002 |
| 417 | 1.873646 | 0.000002 | 1.801567 | 0.000002 |
| 418 | 1.873314 | 0.000002 | 1.801272 | 0.000002 |
| 419 | 1.872985 | 0.000002 | 1.80098 | 0.000002 |
| 420 | 1.872659 | 0.000002 | 1.80069 | 0.000002 |
| 421 | 1.872336 | 0.000002 | 1.800402 | 0.000002 |
| 422 | 1.872015 | 0.000002 | 1.800117 | 0.000002 |
| 423 | 1.871697 | 0.000002 | 1.799834 | 0.000002 |
| 424 | 1.871381 | 0.000002 | 1.799554 | 0.000002 |
| 425 | 1.871068 | 0.000002 | 1.799275 | 0.000002 |
| 426 | 1.870758 | 0.000002 | 1.798999 | 0.000002 |
| 427 | 1.87045 | 0.000002 | 1.798726 | 0.000002 |
| 428 | 1.870144 | 0.000002 | 1.798454 | 0.000002 |
| 429 | 1.869841 | 0.000002 | 1.798185 | 0.000002 |
| 430 | 1.86954 | 0.000002 | 1.797917 | 0.000002 |
| 431 | 1.869242 | 0.000002 | 1.797652 | 0.000002 |
| 432 | 1.868946 | 0.000002 | 1.79739 | 0.000002 |
| 433 | 1.868652 | 0.000002 | 1.797129 | 0.000001 |
| 434 | 1.868361 | 0.000002 | 1.79687 | 0.000001 |
| 435 | 1.868072 | 0.000002 | 1.796613 | 0.000001 |
| 436 | 1.867785 | 0.000002 | 1.796359 | 0.000001 |
| 437 | 1.867501 | 0.000002 | 1.796106 | 0.000001 |
| 438 | 1.867219 | 0.000002 | 1.795855 | 0.000001 |
| 439 | 1.866938 | 0.000002 | 1.795607 | 0.000001 |
| 440 | 1.866661 | 0.000002 | 1.79536 | 0.000001 |
| 441 | 1.866385 | 0.000002 | 1.795115 | 0.000001 |
| 442 | 1.866111 | 0.000002 | 1.794872 | 0.000001 |
| 443 | 1.86584 | 0.000002 | 1.794631 | 0.000001 |
| 444 | 1.86557 | 0.000002 | 1.794392 | 0.000001 |
| 445 | 1.865303 | 0.000002 | 1.794155 | 0.000001 |
| 446 | 1.865037 | 0.000002 | 1.793919 | 0.000001 |
| 447 | 1.864774 | 0.000002 | 1.793685 | 0.000001 |
| 448 | 1.864513 | 0.000003 | 1.793454 | 0.000001 |
| 449 | 1.864253 | 0.000003 | 1.793223 | 0.000001 |
| 450 | 1.863996 | 0.000003 | 1.792995 | 0.000001 |
| 451 | 1.86374 | 0.000003 | 1.792768 | 0.000001 |
| 452 | 1.863487 | 0.000003 | 1.792543 | 0.000001 |
| 453 | 1.863235 | 0.000003 | 1.79232 | 0.000001 |
| 454 | 1.862985 | 0.000003 | 1.792098 | 0.000001 |
| 455 | 1.862737 | 0.000003 | 1.791878 | 0.000001 |
| 456 | 1.862491 | 0.000003 | 1.79166 | 0.000001 |
| 457 | 1.862247 | 0.000003 | 1.791443 | 0.000001 |
| 458 | 1.862004 | 0.000003 | 1.791228 | 0.000001 |
| 459 | 1.861763 | 0.000003 | 1.791015 | 0.000001 |
| 460 | 1.861524 | 0.000003 | 1.790803 | 0.000001 |
| 461 | 1.861287 | 0.000003 | 1.790593 | 0.000001 |
| 462 | 1.861052 | 0.000003 | 1.790384 | 0.000001 |
| 463 | 1.860818 | 0.000003 | 1.790177 | 0.000001 |
| 464 | 1.860586 | 0.000003 | 1.789971 | 0.000001 |
| 465 | 1.860355 | 0.000003 | 1.789767 | 0.000001 |
| 466 | 1.860126 | 0.000003 | 1.789564 | 0.000001 |
| 467 | 1.859899 | 0.000003 | 1.789363 | 0.000001 |
| 468 | 1.859674 | 0.000003 | 1.789163 | 0.000001 |
| 469 | 1.85945 | 0.000003 | 1.788964 | 0.000001 |
| 470 | 1.859228 | 0.000003 | 1.788767 | 0.000001 |
| 471 | 1.859007 | 0.000003 | 1.788572 | 0.000001 |
| 472 | 1.858788 | 0.000003 | 1.788378 | 0.000001 |
| 473 | 1.85857 | 0.000003 | 1.788185 | 0.000001 |
| 474 | 1.858354 | 0.000003 | 1.787993 | 0.000001 |
| 475 | 1.85814 | 0.000003 | 1.787803 | 0.000001 |
| 476 | 1.857927 | 0.000003 | 1.787615 | 0.000001 |
| 477 | 1.857715 | 0.000003 | 1.787427 | 0.000001 |
| 478 | 1.857505 | 0.000003 | 1.787241 | 0.000001 |
| 479 | 1.857296 | 0.000003 | 1.787056 | 0.000001 |
| 480 | 1.857089 | 0.000003 | 1.786873 | 0.000001 |
| 481 | 1.856884 | 0.000003 | 1.786691 | 0.000001 |
| 482 | 1.856679 | 0.000003 | 1.78651 | 0.000001 |
| 483 | 1.856476 | 0.000003 | 1.78633 | 0.000001 |
| 484 | 1.856275 | 0.000003 | 1.786152 | 0.000001 |
| 485 | 1.856075 | 0.000003 | 1.785974 | 0.000001 |
| 486 | 1.855876 | 0.000003 | 1.785798 | 0.000001 |
| 487 | 1.855679 | 0.000003 | 1.785624 | 0.000001 |
| 488 | 1.855483 | 0.000003 | 1.78545 | 0.000001 |
| 489 | 1.855288 | 0.000003 | 1.785278 | 0.000001 |
| 490 | 1.855095 | 0.000003 | 1.785107 | 0.000001 |
| 491 | 1.854903 | 0.000003 | 1.784936 | 0.000001 |
| 492 | 1.854712 | 0.000003 | 1.784768 | 0.000001 |
| 493 | 1.854522 | 0.000003 | sty.46 | 0.000001 |
| 494 | 1.854334 | 0.000003 | 1.784433 | 0.000001 |
| 495 | 1.854147 | 0.000003 | 1.784268 | 0.000001 |
| 496 | 1.853961 | 0.000003 | 1.784103 | 0.000001 |
| 497 | 1.853777 | 0.000003 | 1.78394 | 0.000001 |
| 498 | 1.853594 | 0.000003 | 1.783778 | 0.000001 |
| 499 | 1.853412 | 0.000003 | 1.783617 | 0.000001 |
| 500 | 1.853231 | 0.000003 | 1.783457 | 0.000001 |
| 501 | 1.853051 | 0.000003 | 1.783298 | 0.000001 |
| 502 | 1.852873 | 0.000003 | 1.78314 | 0.000001 |
| 503 | 1.852695 | 0.000003 | 1.782983 | 0.000001 |
| 504 | 1.852519 | 0.000003 | 1.782827 | 0.000001 |
| 505 | 1.852344 | 0.000003 | 1.782672 | 0.000001 |
| 506 | 1.85217 | 0.000003 | 1.782518 | 0.000001 |
| 507 | 1.851998 | 0.000003 | 1.782366 | 0.000001 |
| 508 | 1.851826 | 0.000003 | 1.782214 | 0.000001 |
| 509 | 1.851656 | 0.000003 | 1.782063 | 0.000001 |
| 510 | 1.851486 | 0.000003 | 1.781913 | 0.000001 |
| 511 | 1.851318 | 0.000003 | 1.781764 | 0.000001 |
| 512 | 1.851151 | 0.000003 | 1.781616 | 0.000001 |
| 513 | 1.850985 | 0.000003 | 1.781469 | 0.000001 |
| 514 | 1.85082 | 0.000003 | 1.781323 | 0.000001 |
| 515 | 1.850656 | 0.000003 | 1.781178 | 0.000001 |
| 516 | 1.850493 | 0.000003 | 1.781034 | 0.000001 |
| 517 | 1.850331 | 0.000003 | 1.780891 | 0.000001 |
| 518 | 1.85017 | 0.000003 | 1.780749 | 0.000001 |
| 519 | 1.85001 | 0.000003 | 1.780607 | 0.000001 |
| 520 | 1.849851 | 0.000003 | 1.780467 | 0.000001 |
| 521 | 1.849693 | 0.000003 | 1.780327 | 0.000001 |
| 522 | 1.849536 | 0.000003 | 1.780188 | 0.000001 |
| 523 | 1.84938 | 0.000003 | 1.78005 | 0.000001 |
| 524 | 1.849226 | 0.000003 | 1.779913 | 0.000001 |
| 525 | 1.849072 | 0.000003 | 1.779777 | 0.000001 |
| 526 | 1.848919 | 0.000003 | 1.779642 | 0.000001 |
| 527 | 1.848767 | 0.000003 | 1.779508 | 0.000001 |
| 528 | 1.848616 | 0.000003 | 1.779374 | 0.000001 |
| 529 | 1.848465 | 0.000003 | 1.779241 | 0.000001 |
| 530 | 1.848316 | 0.000003 | 1.779109 | 0.000001 |
| 531 | 1.848168 | 0.000003 | 1.778978 | 0.000001 |
| 532 | 1.84802 | 0.000003 | 1.778848 | 0.000001 |
| 533 | 1.847874 | 0.000003 | 1.778718 | 0.000001 |
| 534 | 1.847728 | 0.000003 | 1.778589 | 0.000001 |
| 535 | 1.847584 | 0.000003 | 1.778461 | 0.000001 |
| 536 | 1.84744 | 0.000003 | 1.778334 | 0.000001 |
| 537 | 1.847297 | 0.000003 | 1.778208 | 0.000001 |
| 538 | 1.847155 | 0.000003 | 1.778082 | 0.000001 |
| 539 | 1.847013 | 0.000003 | 1.777957 | 0.000001 |
| 540 | 1.846873 | 0.000003 | 1.777833 | 0.000001 |
| 541 | 1.846733 | 0.000003 | 1.77771 | 0.000001 |
| 542 | 1.846595 | 0.000003 | 1.777587 | 0.000001 |
| 543 | 1.846457 | 0.000003 | 1.777465 | 0.000001 |
| 544 | 1.84632 | 0.000003 | 1.777344 | 0.000001 |
| 545 | 1.846184 | 0.000003 | 1.777224 | 0.000001 |
| 546 | 1.846048 | 0.000003 | 1.777104 | 0.000001 |
| 547 | 1.845914 | 0.000003 | 1.776985 | 0.000001 |
| 548 | 1.84578 | 0.000004 | 1.776867 | 0.000001 |
| 549 | 1.845647 | 0.000004 | 1.776749 | 0.000001 |
| 550 | 1.845514 | 0.000004 | 1.776632 | 0.000001 |
| 551 | 1.845383 | 0.000004 | 1.776516 | 0.000001 |
| 552 | 1.845252 | 0.000004 | 1.776401 | 0.000001 |
| 553 | 1.845122 | 0.000004 | 1.776286 | 0.000001 |
| 554 | 1.844993 | 0.000004 | 1.776172 | 0.000001 |
| 555 | 1.844865 | 0.000004 | 1.776058 | 0.000001 |
| 556 | 1.844737 | 0.000004 | 1.775945 | 0.000001 |
| 557 | 1.84461 | 0.000004 | 1.775833 | 0.000001 |
| 558 | 1.844484 | 0.000004 | 1.775722 | 0.000001 |
| 559 | 1.844359 | 0.000004 | 1.775611 | 0.000001 |
| 560 | 1.844234 | 0.000004 | 1.775501 | 0.000001 |
| 561 | 1.84411 | 0.000004 | 1.775391 | 0.000001 |
| 562 | 1.843987 | 0.000004 | 1.775282 | 0.000001 |
| 563 | 1.843864 | 0.000004 | 1.775174 | 0.000001 |
| 564 | 1.843742 | 0.000004 | 1.775066 | 0.000001 |
| 565 | 1.843621 | 0.000004 | 1.774959 | 0.000001 |
| 566 | 1.843501 | 0.000004 | 1.774852 | 0.000001 |
| 567 | 1.843381 | 0.000004 | 1.774747 | 0.000001 |
| 568 | 1.843262 | 0.000004 | 1.774641 | 0.000001 |
| 569 | 1.843143 | 0.000004 | 1.774537 | 0.000001 |
| 570 | 1.843026 | 0.000004 | 1.774433 | 0.000001 |
| 571 | 1.842909 | 0.000004 | 1.774329 | 0.000001 |
| 572 | 1.842792 | 0.000004 | 1.774226 | 0.000001 |
| 573 | 1.842676 | 0.000004 | 1.774124 | 0.000001 |
| 574 | 1.842561 | 0.000004 | 1.774022 | 0.000001 |
| 575 | 1.842447 | 0.000004 | 1.773921 | 0.000001 |
| 576 | 1.842333 | 0.000004 | 1.77382 | 0.000001 |
| 577 | 1.84222 | 0.000004 | 1.77372 | 0.000001 |
| 578 | 1.842107 | 0.000004 | 1.773621 | 0.000001 |
| 579 | 1.841995 | 0.000004 | 1.773522 | 0.000001 |
| 580 | 1.841884 | 0.000004 | 1.773424 | 0.000001 |
| 581 | 1.841773 | 0.000004 | 1.773326 | 0.000001 |
| 582 | 1.841663 | 0.000004 | 1.773229 | 0.000001 |
| 583 | 1.841554 | 0.000004 | 1.773132 | 0.000001 |
| 584 | 1.841445 | 0.000004 | 1.773036 | 0.000001 |
| 585 | 1.841336 | 0.000004 | 1.77294 | 0.000001 |
| 586 | 1.841229 | 0.000004 | 1.772845 | 0.000001 |
| 587 | 1.841122 | 0.000004 | 1.77275 | 0.000001 |
| 588 | 1.841015 | 0.000004 | 1.772656 | 0.000001 |
| 589 | 1.840909 | 0.000004 | 1.772562 | 0.000001 |
| 590 | 1.840804 | 0.000004 | 1.772469 | 0.000001 |
| 591 | 1.840699 | 0.000004 | 1.772377 | 0.000001 |
| 592 | 1.840595 | 0.000004 | 1.772285 | 0.000001 |
| 593 | 1.840491 | 0.000004 | 1.772193 | 0.000001 |
| 594 | 1.840388 | 0.000004 | 1.772102 | 0.000001 |
| 595 | 1.840286 | 0.000004 | 1.772011 | 0.000001 |
| 596 | 1.840184 | 0.000004 | 1.771921 | 0.000001 |
| 597 | 1.840082 | 0.000004 | 1.771832 | 0.000001 |
| 598 | 1.839981 | 0.000004 | 1.771742 | 0.000001 |
| 599 | 1.839881 | 0.000004 | 1.771654 | 0.000001 |
| 600 | 1.839781 | 0.000004 | 1.771566 | 0.000001 |
| 601 | 1.839682 | 0.000004 | 1.771478 | 0.000001 |
| 602 | 1.839583 | 0.000004 | 1.771391 | 0.000001 |
| 603 | 1.839485 | 0.000004 | 1.771304 | 0.000001 |
| 604 | 1.839387 | 0.000004 | 1.771218 | 0.000001 |
| 605 | 1.83929 | 0.000004 | 1.771132 | 0.000001 |
| 606 | 1.839194 | 0.000004 | 1.771046 | 0.000001 |
| 607 | 1.839097 | 0.000004 | 1.770961 | 0.000001 |
| 608 | 1.839002 | 0.000004 | 1.770877 | 0.000001 |
| 609 | 1.838907 | 0.000004 | 1.770793 | 0.000001 |
| 610 | 1.838812 | 0.000004 | 1.770709 | 0.000001 |
| 611 | 1.838718 | 0.000004 | 1.770626 | 0.000001 |
| 612 | 1.838624 | 0.000004 | 1.770543 | 0.000001 |
| 613 | 1.838531 | 0.000004 | 1.770461 | 0.000001 |
| 614 | 1.838438 | 0.000004 | 1.770379 | 0.000001 |
| 615 | 1.838346 | 0.000004 | 1.770297 | 0.000001 |
| 616 | 1.838254 | 0.000004 | 1.770216 | 0.000001 |
| 617 | 1.838163 | 0.000004 | 1.770136 | 0.000001 |
| 618 | 1.838072 | 0.000004 | 1.770055 | 0.000001 |
| 619 | 1.837982 | 0.000004 | 1.769976 | 0.000001 |
| 620 | 1.837892 | 0.000004 | 1.769896 | 0.000001 |
| 621 | 1.837803 | 0.000004 | 1.769817 | 0.000001 |
| 622 | 1.837714 | 0.000004 | 1.769739 | 0.000001 |
| 623 | 1.837626 | 0.000004 | 1.76966 | 0.000001 |
| 624 | 1.837537 | 0.000004 | 1.769583 | 0.000001 |
| 625 | 1.83745 | 0.000004 | 1.769505 | 0.000001 |
| 626 | 1.837363 | 0.000004 | 1.769428 | 0.000001 |
| 627 | 1.837276 | 0.000004 | 1.769352 | 0.000001 |
| 628 | 1.83719 | 0.000004 | 1.769276 | 0.000001 |
| 629 | 1.837104 | 0.000004 | sty.92 | 0.000001 |
| 630 | 1.837019 | 0.000004 | 1.769124 | 0.000001 |
| 631 | 1.836934 | 0.000004 | 1.769049 | 0.000001 |
| 632 | 1.836849 | 0.000004 | 1.768974 | 0.000001 |
| 633 | 1.836765 | 0.000004 | sty.89 | 0.000001 |
| 634 | 1.836681 | 0.000005 | 1.768826 | 0.000001 |
| 635 | 1.836598 | 0.000005 | 1.768753 | 0.000001 |
| 636 | 1.836516 | 0.000005 | 1.768679 | 0.000001 |
| 637 | 1.836433 | 0.000005 | 1.768607 | 0.000001 |
| 638 | 1.836351 | 0.000005 | 1.768534 | 0.000001 |
| 639 | 1.836269 | 0.000005 | 1.768462 | 0.000001 |
| 640 | 1.836188 | 0.000005 | 1.76839 | 0.000001 |
| 641 | 1.836107 | 0.000005 | 1.768319 | 0.000001 |
| 642 | 1.836027 | 0.000005 | 1.768248 | 0.000001 |
| 643 | 1.835947 | 0.000005 | 1.768177 | 0.000001 |
| 644 | 1.835868 | 0.000005 | 1.768107 | 0.000001 |
| 645 | 1.835788 | 0.000005 | 1.768037 | 0.000001 |
| 646 | 1.83571 | 0.000005 | 1.767967 | 0.000001 |
| 647 | 1.835631 | 0.000005 | 1.767898 | 0.000001 |
| 648 | 1.835553 | 0.000005 | 1.767829 | 0.000001 |
| 649 | 1.835475 | 0.000005 | 1.76776 | 0.000001 |
| 650 | 1.835398 | 0.000005 | 1.767692 | 0.000001 |
| 651 | 1.835321 | 0.000005 | 1.767624 | 0.000001 |
| 652 | 1.835245 | 0.000005 | 1.767556 | 0.000001 |
| 653 | 1.835169 | 0.000005 | 1.767489 | 0.000001 |
| 654 | 1.835093 | 0.000005 | 1.767422 | 0.000001 |
| 655 | 1.835017 | 0.000005 | 1.767355 | 0.000001 |
| 656 | 1.834942 | 0.000005 | 1.767289 | 0.000001 |
| 657 | 1.834868 | 0.000005 | 1.767223 | 0.000001 |
| 658 | 1.834793 | 0.000005 | 1.767157 | 0.000001 |
| 659 | 1.834719 | 0.000005 | 1.767092 | 0.000001 |
| 660 | 1.834646 | 0.000005 | 1.767027 | 0.000001 |
| 661 | 1.834573 | 0.000005 | 1.766962 | 0.000001 |
| 662 | sty.45 | 0.000005 | 1.766897 | 0.000001 |
| 663 | 1.834427 | 0.000005 | 1.766833 | 0.000001 |
| 664 | 1.834355 | 0.000005 | 1.766769 | 0.000001 |
| 665 | 1.834283 | 0.000005 | 1.766706 | 0.000001 |
| 666 | 1.834211 | 0.000005 | 1.766642 | 0.000001 |
| 667 | 1.83414 | 0.000005 | 1.76658 | 0.000001 |
| 668 | 1.834069 | 0.000005 | 1.766517 | 0.000001 |
| 669 | 1.833999 | 0.000005 | 1.766454 | 0.000001 |
| 670 | 1.833929 | 0.000005 | 1.766392 | 0.000001 |
| 671 | 1.833859 | 0.000005 | 1.766331 | 0.000001 |
| 672 | 1.833789 | 0.000005 | 1.766269 | 0.000001 |
| 673 | 1.83372 | 0.000005 | 1.766208 | 0.000001 |
| 674 | 1.833651 | 0.000005 | 1.766147 | 0.000001 |
| 675 | 1.833583 | 0.000005 | 1.766086 | 0.000001 |
| 676 | 1.833515 | 0.000005 | 1.766026 | 0.000001 |
| 677 | 1.833447 | 0.000005 | 1.765966 | 0.000001 |
| 678 | 1.833379 | 0.000005 | 1.765906 | 0.000001 |
| 679 | 1.833312 | 0.000005 | 1.765847 | 0.000001 |
| 680 | 1.833245 | 0.000005 | 1.765787 | 0.000001 |
| 681 | 1.833178 | 0.000005 | 1.765728 | 0.000001 |
| 682 | 1.833112 | 0.000005 | 1.76567 | 0.000001 |
| 683 | 1.833046 | 0.000005 | 1.765611 | 0.000001 |
| 684 | 1.83298 | 0.000005 | 1.765553 | 0.000001 |
| 685 | 1.832915 | 0.000005 | 1.765495 | 0.000001 |
| 686 | 1.83285 | 0.000005 | 1.765438 | 0.000001 |
| 687 | 1.832785 | 0.000005 | 1.76538 | 0.000001 |
| 688 | 1.83272 | 0.000005 | 1.765323 | 0.000001 |
| 689 | 1.832656 | 0.000005 | 1.765266 | 0.000001 |
| 690 | 1.832592 | 0.000005 | 1.76521 | 0.000001 |
| 691 | 1.832528 | 0.000005 | 1.765154 | 0.000001 |
| 692 | 1.832465 | 0.000005 | 1.765097 | 0.000001 |
| 693 | 1.832402 | 0.000005 | 1.765042 | 0.000001 |
| 694 | 1.832339 | 0.000005 | 1.764986 | 0.000001 |
| 695 | 1.832277 | 0.000005 | 1.764931 | 0.000001 |
| 696 | 1.832214 | 0.000005 | 1.764876 | 0.000001 |
| 697 | 1.832152 | 0.000005 | 1.764821 | 0.000001 |
| 698 | 1.832091 | 0.000005 | 1.764766 | 0.000001 |
| 699 | 1.832029 | 0.000005 | 1.764712 | 0.000001 |
| 700 | 1.831968 | 0.000005 | 1.764658 | 0.000001 |
| 701 | 1.831907 | 0.000005 | 1.764604 | 0.000001 |
| 702 | 1.831847 | 0.000005 | 1.76455 | 0.000001 |
| 703 | 1.831787 | 0.000005 | 1.764497 | 0.000001 |
| 704 | 1.831727 | 0.000005 | 1.764444 | 0.000001 |
| 705 | 1.831667 | 0.000005 | 1.764391 | 0.000001 |
| 706 | 1.831607 | 0.000005 | 1.764338 | 0.000001 |
| 707 | 1.831548 | 0.000005 | 1.764286 | 0.000001 |
| 708 | 1.831489 | 0.000005 | 1.764234 | 0.000001 |
| 709 | 1.83143 | 0.000005 | 1.764182 | 0.000001 |
| 710 | 1.831372 | 0.000005 | 1.76413 | 0.000001 |
| 711 | 1.831314 | 0.000005 | 1.764078 | 0.000001 |
| 712 | 1.831256 | 0.000005 | 1.764027 | 0.000001 |
| 713 | 1.831198 | 0.000005 | 1.763976 | 0.000001 |
| 714 | 1.831141 | 0.000005 | 1.763925 | 0.000001 |
| 715 | 1.831084 | 0.000005 | 1.763875 | 0.000001 |
| 716 | 1.831027 | 0.000005 | 1.763824 | 0.000001 |
| 717 | 1.83097 | 0.000005 | 1.763774 | 0.000001 |
| 718 | 1.830914 | 0.000005 | 1.763724 | 0.000001 |
| 719 | 1.830858 | 0.000005 | 1.763674 | 0.000001 |
| 720 | 1.830802 | 0.000005 | 1.763625 | 0.000001 |
| 721 | 1.830746 | 0.000005 | 1.763576 | 0.000001 |
| 722 | 1.830691 | 0.000005 | 1.763527 | 0.000001 |
| 723 | 1.830635 | 0.000005 | 1.763478 | 0.000001 |
| 724 | 1.83058 | 0.000005 | 1.763429 | 0.000001 |
| 725 | 1.830526 | 0.000005 | 1.763381 | 0.000001 |
| 726 | 1.830471 | 0.000005 | 1.763332 | 0.000001 |
| 727 | 1.830417 | 0.000005 | 1.763284 | 0.000001 |
| 728 | 1.830363 | 0.000005 | 1.763236 | 0.000001 |
| 729 | 1.830309 | 0.000005 | 1.763189 | 0.000001 |
| 730 | 1.830256 | 0.000005 | 1.763141 | 0.000001 |
| 731 | 1.830202 | 0.000005 | 1.763094 | 0.000001 |

| 732 | 1.830149 | 0.000005 | 1.763047 | 0.000001 |
|---:|---:|---:|---:|---:|
| 733 | 1.830096 | 0.000005 | 1.763 | 0.000001 |
| 734 | 1.830044 | 0.000005 | 1.762954 | 0.000001 |
| 735 | 1.829991 | 0.000005 | 1.762907 | 0.000001 |
| 736 | 1.829939 | 0.000005 | 1.762861 | 0.000001 |
| 737 | 1.829887 | 0.000006 | 1.762815 | 0.000001 |
| 738 | 1.829836 | 0.000006 | 1.762769 | 0.000001 |
| 739 | 1.829784 | 0.000006 | 1.762723 | 0.000001 |
| 740 | 1.829733 | 0.000006 | 1.762678 | 0.000001 |
| 741 | 1.829682 | 0.000006 | 1.762632 | 0.000001 |
| 742 | 1.829631 | 0.000006 | 1.762587 | 0.000001 |
| 743 | 1.82958 | 0.000006 | 1.762542 | 0.000001 |
| 744 | 1.82953 | 0.000006 | 1.762498 | 0.000001 |
| 745 | 1.829479 | 0.000006 | 1.762453 | 0.000001 |
| 746 | 1.829429 | 0.000006 | 1.762409 | 0.000001 |
| 747 | 1.82938 | 0.000006 | 1.762365 | 0.000001 |
| 748 | 1.82933 | 0.000006 | 1.762321 | 0.000001 |
| 749 | 1.829281 | 0.000006 | 1.762277 | 0.000001 |
| 750 | 1.829232 | 0.000006 | 1.762233 | 0.000001 |
| 751 | 1.829183 | 0.000006 | 1.76219 | 0.000001 |
| 752 | 1.829134 | 0.000006 | 1.762147 | 0.000001 |
| 753 | 1.829085 | 0.000006 | 1.762104 | 0.000001 |
| 754 | 1.829037 | 0.000006 | 1.762061 | 0.000001 |
| 755 | 1.828989 | 0.000006 | 1.762018 | 0.000001 |
| 756 | 1.828941 | 0.000006 | 1.761975 | 0.000001 |
| 757 | 1.828893 | 0.000006 | 1.761933 | 0.000001 |
| 758 | 1.828845 | 0.000006 | 1.761891 | 0.000001 |
| 759 | 1.828798 | 0.000006 | 1.761849 | 0.000001 |
| 760 | 1.828751 | 0.000006 | 1.761807 | 0.000001 |
| 761 | 1.828704 | 0.000006 | 1.761765 | 0.000001 |
| 762 | 1.828657 | 0.000006 | 1.761724 | 0.000001 |
| 763 | 1.82861 | 0.000006 | 1.761682 | 0.000001 |
| 764 | 1.828564 | 0.000006 | 1.761641 | 0.000001 |
| 765 | 1.828518 | 0.000006 | sty.16 | 0.000001 |
| 766 | 1.828472 | 0.000006 | 1.761559 | 0.000001 |
| 767 | 1.828426 | 0.000006 | 1.761519 | 0.000001 |
| 768 | 1.82838 | 0.000006 | 1.761478 | 0.000001 |
| 769 | 1.828335 | 0.000006 | 1.761438 | 0.000001 |
| 770 | 1.828289 | 0.000006 | 1.761397 | 0.000001 |
| 771 | 1.828244 | 0.000006 | 1.761357 | 0.000001 |
| 772 | 1.828199 | 0.000006 | 1.761318 | 0.000001 |
| 773 | 1.828154 | 0.000006 | 1.761278 | 0.000001 |
| 774 | 1.82811 | 0.000006 | 1.761238 | 0.000001 |
| 775 | 1.828066 | 0.000006 | 1.761199 | 0.000001 |
| 776 | 1.828021 | 0.000006 | 1.76116 | 0.000001 |
| 777 | 1.827977 | 0.000006 | 1.761121 | 0.000001 |
| 778 | 1.827933 | 0.000006 | 1.761082 | 0.000001 |
| 779 | 1.82789 | 0.000006 | 1.761043 | 0.000001 |
| 780 | 1.827846 | 0.000006 | 1.761004 | 0.000001 |
| 781 | 1.827803 | 0.000006 | 1.760966 | 0.000001 |
| 782 | 1.82776 | 0.000006 | 1.760927 | 0.000001 |
| 783 | 1.827717 | 0.000006 | 1.760889 | 0.000001 |
| 784 | 1.827674 | 0.000006 | 1.760851 | 0.000001 |
| 785 | 1.827631 | 0.000006 | 1.760813 | 0.000001 |
| 786 | 1.827589 | 0.000006 | 1.760776 | 0.000001 |
| 787 | 1.827546 | 0.000006 | 1.760738 | 0.000001 |
| 788 | 1.827504 | 0.000006 | sty.07 | 0.000001 |
| 789 | 1.827462 | 0.000006 | 1.760663 | 0.000001 |
| 790 | 1.827421 | 0.000006 | 1.760626 | 0.000001 |
| 791 | 1.827379 | 0.000006 | 1.760589 | 0.000001 |
| 792 | 1.827338 | 0.000006 | 1.760552 | 0.000001 |
| 793 | 1.827296 | 0.000006 | 1.760515 | 0.000001 |
| 794 | 1.827255 | 0.000006 | 1.760479 | 0.000001 |
| 795 | 1.827214 | 0.000006 | 1.760442 | 0.000001 |
| 796 | 1.827173 | 0.000006 | 1.760406 | 0.000001 |
| 797 | 1.827132 | 0.000006 | 1.76037 | 0.000001 |
| 798 | 1.827092 | 0.000006 | 1.760334 | 0.000001 |
| 799 | 1.827052 | 0.000006 | 1.760298 | 0.000001 |
| 800 | 1.827011 | 0.000006 | 1.760262 | 0.000001 |
| 801 | 1.826971 | 0.000006 | 1.760227 | 0.000001 |
| 802 | 1.826931 | 0.000006 | 1.760191 | 0.000001 |
| 803 | 1.826892 | 0.000006 | 1.760156 | 0.000001 |
| 804 | 1.826852 | 0.000006 | 1.760121 | 0.000001 |
| 805 | 1.826813 | 0.000006 | 1.760086 | 0.000001 |
| 806 | 1.826774 | 0.000006 | 1.760051 | 0.000001 |
| 807 | 1.826734 | 0.000006 | 1.760016 | 0.000001 |
| 808 | 1.826695 | 0.000006 | 1.759981 | 0.000001 |
| 809 | 1.826657 | 0.000006 | 1.759947 | 0.000001 |
| 810 | 1.826618 | 0.000006 | 1.759912 | 0.000001 |
| 811 | 1.826579 | 0.000006 | 1.759878 | 0.000001 |
| 812 | 1.826541 | 0.000006 | 1.759844 | 0.000001 |
| 813 | 1.826503 | 0.000006 | 1.75981 | 0.000001 |
| 814 | 1.826465 | 0.000006 | 1.759776 | 0.000001 |
| 815 | 1.826427 | 0.000006 | 1.759742 | 0.000001 |
| 816 | 1.826389 | 0.000006 | 1.759709 | 0.000001 |
| 817 | 1.826351 | 0.000006 | 1.759675 | 0.000001 |
| 818 | 1.826314 | 0.000006 | 1.759642 | 0.000001 |
| 819 | 1.826276 | 0.000006 | 1.759609 | 0.000001 |
| 820 | 1.826239 | 0.000006 | 1.759575 | 0.000001 |
| 821 | 1.826202 | 0.000006 | 1.759542 | 0.000001 |
| 822 | 1.826165 | 0.000006 | 1.759509 | 0.000001 |
| 823 | 1.826128 | 0.000006 | 1.759477 | 0.000001 |
| 824 | 1.826092 | 0.000006 | 1.759444 | 0.000001 |
| 825 | 1.826055 | 0.000006 | 1.759412 | 0.000001 |
| 826 | 1.826019 | 0.000006 | 1.759379 | 0.000001 |
| 827 | 1.825983 | 0.000006 | 1.759347 | 0.000001 |
| 828 | 1.825946 | 0.000006 | 1.759315 | 0.000001 |
| 829 | 1.82591 | 0.000006 | 1.759283 | 0.000001 |
| 830 | 1.825875 | 0.000006 | 1.759251 | 0.000001 |
| 831 | 1.825839 | 0.000006 | 1.759219 | 0.000001 |
| 832 | 1.825803 | 0.000006 | 1.759187 | 0.000001 |
| 833 | 1.825768 | 0.000006 | 1.759156 | 0.000001 |
| 834 | 1.825733 | 0.000006 | 1.759124 | 0.000001 |
| 835 | 1.825697 | 0.000006 | 1.759093 | 0.000001 |
| 836 | 1.825662 | 0.000006 | 1.759062 | 0.000001 |

| 837 | 1.825627 | 0.000006 | 1.75903 | 0.000001 |
|---|---|---|---|---|
| 838 | 1.825593 | 0.000006 | 1.758999 | 0.000001 |
| 839 | 1.825558 | 0.000006 | 1.758969 | 0.000001 |
| 840 | 1.825523 | 0.000006 | 1.758938 | 0.000001 |
| 841 | 1.825489 | 0.000006 | 1.758907 | 0.000001 |
| 842 | 1.825455 | 0.000006 | 1.758877 | 0.000001 |
| 843 | 1.82542 | 0.000006 | 1.758846 | 0.000001 |
| 844 | 1.825387 | 0.000006 | 1.758816 | 0.000001 |
| 845 | 1.825353 | 0.000006 | 1.758785 | 0.000001 |
| 846 | 1.825319 | 0.000006 | 1.758755 | 0.000001 |
| 847 | 1.825285 | 0.000006 | 1.758725 | 0.000001 |
| 848 | 1.825252 | 0.000006 | 1.758695 | 0.000001 |
| 849 | 1.825218 | 0.000006 | 1.758666 | 0.000001 |
| 850 | 1.825185 | 0.000006 | 1.758636 | 0.000001 |
| 851 | 1.825152 | 0.000006 | 1.758606 | 0.000001 |
| 852 | 1.825119 | 0.000006 | 1.758577 | 0.000001 |
| 853 | 1.825086 | 0.000006 | 1.758548 | 0.000001 |
| 854 | 1.825053 | 0.000006 | 1.758518 | 0.000001 |
| 855 | 1.82502 | 0.000006 | 1.758489 | 0.000001 |
| 856 | 1.824988 | 0.000006 | 1.75846 | 0.000001 |
| 857 | 1.824955 | 0.000006 | 1.758431 | 0.000001 |
| 858 | 1.824923 | 0.000006 | 1.758402 | 0.000001 |
| 859 | 1.824891 | 0.000006 | 1.758373 | 0.000001 |
| 860 | 1.824859 | 0.000006 | 1.758345 | 0.000001 |
| 861 | 1.824827 | 0.000006 | 1.758316 | 0.000001 |
| 862 | 1.824795 | 0.000006 | 1.758288 | 0.000001 |
| 863 | 1.824763 | 0.000006 | 1.758259 | 0.000001 |
| 864 | 1.824731 | 0.000006 | 1.758231 | 0.000001 |
| 865 | sty.47 | 0.000006 | 1.758203 | 0.000001 |
| 866 | 1.824668 | 0.000006 | 1.758175 | 0.000001 |
| 867 | 1.824637 | 0.000006 | 1.758147 | 0.000001 |
| 868 | 1.824606 | 0.000006 | 1.758119 | 0.000001 |
| 869 | 1.824575 | 0.000006 | 1.758091 | 0.000001 |
| 870 | 1.824544 | 0.000006 | 1.758064 | 0.000001 |
| 871 | 1.824513 | 0.000006 | 1.758036 | 0.000001 |
| 872 | 1.824482 | 0.000006 | 1.758009 | 0.000001 |
| 873 | 1.824452 | 0.000006 | 1.757981 | 0.000001 |
| 874 | 1.824421 | 0.000006 | 1.757954 | 0.000001 |
| 875 | 1.824391 | 0.000006 | 1.757927 | 0.000001 |
| 876 | 1.82436 | 0.000006 | sty.79 | 0.000001 |
| 877 | 1.82433 | 0.000006 | 1.757873 | 0.000001 |
| 878 | sty.43 | 0.000006 | 1.757846 | 0.000001 |
| 879 | 1.82427 | 0.000006 | 1.757819 | 0.000001 |
| 880 | 1.82424 | 0.000006 | 1.757792 | 0.000001 |
| 881 | 1.82421 | 0.000006 | 1.757766 | 0.000001 |
| 882 | 1.824181 | 0.000006 | 1.757739 | 0.000001 |
| 883 | 1.824151 | 0.000006 | 1.757713 | 0.000001 |
| 884 | 1.824122 | 0.000006 | 1.757686 | 0.000001 |
| 885 | 1.824092 | 0.000006 | 1.75766 | 0.000001 |
| 886 | 1.824063 | 0.000006 | 1.757634 | 0.000001 |
| 887 | 1.824034 | 0.000006 | 1.757608 | 0.000001 |
| 888 | 1.824005 | 0.000006 | 1.757582 | 0.000001 |
| 889 | 1.823976 | 0.000006 | 1.757556 | 0.000001 |
| 890 | 1.823947 | 0.000006 | 1.75753 | 0.000001 |
| 891 | 1.823918 | 0.000006 | 1.757504 | 0.000001 |
| 892 | 1.823889 | 0.000006 | 1.757478 | 0.000001 |
| 893 | 1.823861 | 0.000006 | 1.757453 | 0.000001 |
| 894 | 1.823832 | 0.000006 | 1.757427 | 0.000001 |
| 895 | 1.823804 | 0.000006 | 1.757402 | 0.000001 |
| 896 | 1.823776 | 0.000006 | 1.757377 | 0.000001 |
| 897 | 1.823748 | 0.000006 | 1.757352 | 0.000001 |
| 898 | 1.82372 | 0.000006 | 1.757326 | 0.000001 |
| 899 | 1.823691 | 0.000006 | 1.757301 | 0.000001 |
| 900 | 1.823664 | 0.000006 | 1.757276 | 0.000001 |
| 901 | 1.823636 | 0.000006 | 1.757252 | 0.000001 |
| 902 | 1.823608 | 0.000006 | 1.757227 | 0.000001 |
| 903 | 1.823581 | 0.000006 | 1.757202 | 0.000001 |
| 904 | 1.823553 | 0.000006 | 1.757177 | 0.000001 |
| 905 | 1.823526 | 0.000006 | 1.757153 | 0.000001 |
| 906 | 1.823498 | 0.000006 | 1.757128 | 0.000001 |
| 907 | 1.823471 | 0.000006 | 1.757104 | 0.000001 |
| 908 | 1.823444 | 0.000006 | 1.757079 | 0.000001 |
| 909 | 1.823417 | 0.000006 | 1.757055 | 0.000001 |
| 910 | 1.82339 | 0.000006 | 1.757031 | 0.000001 |
| 911 | 1.823363 | 0.000006 | 1.757007 | 0.000001 |
| 912 | 1.823336 | 0.000006 | 1.756983 | 0 |
| 913 | 1.82331 | 0.000006 | 1.756959 | 0 |
| 914 | 1.823283 | 0.000006 | 1.756935 | 0 |
| 915 | 1.823256 | 0.000006 | 1.756912 | 0 |
| 916 | 1.82323 | 0.000006 | 1.756888 | 0 |
| 917 | 1.823204 | 0.000006 | 1.756864 | 0 |
| 918 | 1.823178 | 0.000006 | 1.756841 | 0 |
| 919 | 1.823151 | 0.000006 | 1.756817 | 0 |
| 920 | 1.823125 | 0.000006 | 1.756794 | 0 |
| 921 | 1.823099 | 0.000006 | 1.75677 | 0 |
| 922 | 1.823074 | 0.000006 | 1.756747 | 0 |
| 923 | 1.823048 | 0.000006 | 1.756724 | 0 |
| 924 | 1.823022 | 0.000006 | 1.756701 | 0 |
| 925 | 1.822996 | 0.000006 | 1.756678 | 0 |
| 926 | 1.822971 | 0.000006 | 1.756655 | 0 |
| 927 | 1.822945 | 0.000006 | 1.756632 | 0 |
| 928 | 1.82292 | 0.000007 | 1.75661 | 0 |
| 929 | 1.822895 | 0.000007 | 1.756587 | 0 |
| 930 | 1.82287 | 0.000007 | 1.756564 | 0 |
| 931 | 1.822845 | 0.000007 | 1.756542 | 0 |
| 932 | 1.822819 | 0.000007 | 1.756519 | 0 |
| 933 | 1.822794 | 0.000007 | 1.756497 | 0 |
| 934 | 1.82277 | 0.000007 | 1.756474 | 0 |
| 935 | 1.822745 | 0.000007 | 1.756452 | 0 |
| 936 | 1.82272 | 0.000007 | 1.75643 | 0 |
| 937 | 1.822695 | 0.000007 | 1.756408 | 0 |
| 938 | 1.822671 | 0.000007 | 1.756386 | 0 |
| 939 | 1.822646 | 0.000007 | 1.756364 | 0 |
| 940 | 1.822622 | 0.000007 | 1.756342 | 0 |
| 941 | 1.822598 | 0.000007 | 1.75632 | 0 |
| 942 | 1.822574 | 0.000007 | 1.756298 | 0 |
| 943 | 1.822549 | 0.000007 | 1.756276 | 0 |
| 944 | 1.822525 | 0.000007 | 1.756255 | 0 |
| 945 | 1.822501 | 0.000007 | 1.756233 | 0 |
| 946 | 1.822477 | 0.000007 | 1.756212 | 0 |
| 947 | 1.822454 | 0.000007 | 1.75619 | 0 |
| 948 | 1.82243 | 0.000007 | 1.756169 | 0 |
| 949 | 1.822406 | 0.000007 | 1.756148 | 0 |
| 950 | 1.822383 | 0.000007 | 1.756126 | 0 |
| 951 | 1.822359 | 0.000007 | 1.756105 | 0 |
| 952 | 1.822336 | 0.000007 | 1.756084 | 0 |
| 953 | 1.822312 | 0.000007 | 1.756063 | 0 |
| 954 | 1.822289 | 0.000007 | 1.756042 | 0 |
| 955 | 1.822266 | 0.000007 | 1.756021 | 0 |
| 956 | 1.822243 | 0.000007 | 1.756 | 0 |
| 957 | 1.82222 | 0.000007 | 1.75598 | 0 |
| 958 | 1.822197 | 0.000007 | 1.755959 | 0 |
| 959 | 1.822174 | 0.000007 | 1.755938 | 0 |
| 960 | 1.822151 | 0.000007 | 1.755918 | 0 |
| 961 | 1.822128 | 0.000007 | 1.755897 | 0 |
| 962 | 1.822105 | 0.000007 | 1.755877 | 0 |
| 963 | 1.822083 | 0.000007 | 1.755856 | 0 |
| 964 | 1.82206 | 0.000007 | 1.755836 | 0 |
| 965 | 1.822038 | 0.000007 | 1.755816 | 0 |
| 966 | 1.822015 | 0.000007 | 1.755795 | 0 |
| 967 | 1.821993 | 0.000007 | 1.755775 | 0 |
| 968 | 1.821971 | 0.000007 | 1.755755 | 0 |
| 969 | 1.821948 | 0.000007 | 1.755735 | 0 |
| 970 | 1.821926 | 0.000007 | 1.755715 | 0 |
| 971 | 1.821904 | 0.000007 | 1.755695 | 0 |
| 972 | 1.821882 | 0.000007 | 1.755675 | 0 |
| 973 | 1.82186 | 0.000007 | 1.755656 | 0 |
| 974 | 1.821838 | 0.000007 | 1.755636 | 0 |
| 975 | 1.821817 | 0.000007 | 1.755616 | 0 |
| 976 | 1.821795 | 0.000007 | 1.755597 | 0 |
| 977 | 1.821773 | 0.000007 | 1.755577 | 0 |
| 978 | 1.821752 | 0.000007 | 1.755558 | 0 |
| 979 | 1.82173 | 0.000007 | 1.755538 | 0 |
| 980 | 1.821709 | 0.000007 | 1.755519 | 0 |
| 981 | 1.821687 | 0.000007 | 1.755499 | 0 |
| 982 | 1.821666 | 0.000007 | 1.75548 | 0 |
| 983 | 1.821645 | 0.000007 | 1.755461 | 0 |
| 984 | 1.821624 | 0.000007 | 1.755442 | 0 |
| 985 | 1.821603 | 0.000007 | 1.755423 | 0 |
| 986 | 1.821582 | 0.000007 | 1.755404 | 0 |
| 987 | 1.821561 | 0.000007 | 1.755385 | 0 |
| 988 | 1.82154 | 0.000007 | 1.755366 | 0 |
| 989 | 1.821519 | 0.000007 | 1.755347 | 0 |
| 990 | 1.821498 | 0.000007 | 1.755328 | 0 |
| 991 | 1.821477 | 0.000007 | 1.75531 | 0 |
| 992 | 1.821457 | 0.000007 | 1.755291 | 0 |
| 993 | 1.821436 | 0.000007 | 1.755272 | 0 |
| 994 | 1.821416 | 0.000007 | 1.755254 | 0 |
| 995 | 1.821395 | 0.000007 | 1.755235 | 0 |
| 996 | 1.821375 | 0.000007 | 1.755217 | 0 |
| 997 | 1.821354 | 0.000007 | 1.755198 | 0 |
| 998 | 1.821334 | 0.000007 | 1.75518 | 0 |
| 999 | 1.821314 | 0.000007 | 1.755162 | 0 |
| 1000 | 1.821294 | 0.000007 | 1.755143 | 0 |
| 1002.5 | 1.821244 | 0.000007 | 1.755098 | 0 |
| 1005 | 1.821194 | 0.000007 | 1.755053 | 0 |
| 1007.5 | 1.821145 | 0.000007 | 1.755008 | 0 |
| 1010 | 1.821096 | 0.000007 | 1.754964 | 0 |
| 1012.5 | 1.821047 | 0.000007 | 1.75492 | 0 |
| 1015 | 1.820999 | 0.000007 | 1.754876 | 0 |
| 1017.5 | 1.820951 | 0.000007 | 1.754833 | 0 |
| 1020 | 1.820903 | 0.000007 | 1.75479 | 0 |
| 1022.5 | 1.820856 | 0.000007 | 1.754747 | 0 |
| 1025 | 1.820809 | 0.000007 | 1.754704 | 0 |
| 1027.5 | 1.820763 | 0.000007 | 1.754662 | 0 |
| 1030 | 1.820717 | 0.000007 | 1.75462 | 0 |
| 1032.5 | 1.820671 | 0.000007 | 1.754578 | 0 |
| 1035 | 1.820626 | 0.000007 | 1.754537 | 0 |
| 1037.5 | 1.820581 | 0.000007 | 1.754496 | 0 |
| 1040 | 1.820536 | 0.000007 | 1.754455 | 0 |
| 1042.5 | 1.820491 | 0.000007 | 1.754415 | 0 |
| 1045 | 1.820447 | 0.000007 | 1.754375 | 0 |
| 1047.5 | 1.820404 | 0.000007 | 1.754335 | 0 |
| 1050 | 1.82036 | 0.000007 | 1.754295 | 0 |
| 1052.5 | 1.820317 | 0.000007 | 1.754256 | 0 |
| 1055 | 1.820274 | 0.000007 | 1.754217 | 0 |
| 1057.5 | 1.820232 | 0.000007 | 1.754178 | 0 |
| 1060 | 1.820189 | 0.000007 | 1.75414 | 0 |
| 1062.5 | 1.820148 | 0.000007 | 1.754101 | 0 |
| 1065 | 1.820106 | 0.000007 | 1.754063 | 0 |
| 1067.5 | 1.820065 | 0.000007 | 1.754026 | 0 |
| 1070 | 1.820024 | 0.000007 | 1.753988 | 0 |
| 1072.5 | 1.819983 | 0.000007 | 1.753951 | 0 |
| 1075 | 1.819942 | 0.000007 | 1.753914 | 0 |
| 1077.5 | 1.819902 | 0.000007 | 1.753877 | 0 |
| 1080 | 1.819862 | 0.000007 | 1.753841 | 0 |
| 1082.5 | 1.819823 | 0.000007 | 1.753804 | 0 |
| 1085 | 1.819783 | 0.000007 | 1.753768 | 0 |
| 1087.5 | 1.819744 | 0.000007 | 1.753733 | 0 |
| 1090 | 1.819706 | 0.000007 | 1.753697 | 0 |
| 1092.5 | 1.819667 | 0.000007 | 1.753662 | 0 |
| 1095 | 1.819629 | 0.000007 | 1.753627 | 0 |
| 1097.5 | 1.819591 | 0.000007 | 1.753592 | 0 |
| 1100 | 1.819553 | 0.000007 | 1.753557 | 0 |
| 1102.5 | 1.819516 | 0.000007 | 1.753523 | 0 |
| 1105 | 1.819479 | 0.000007 | 1.753489 | 0 |
| 1107.5 | 1.819442 | 0.000007 | 1.753455 | 0 |
| 1110 | 1.819405 | 0.000007 | 1.753421 | 0 |
| 1112.5 | 1.819369 | 0.000007 | 1.753387 | 0 |
| 1115 | 1.819333 | 0.000007 | 1.753354 | 0 |
| 1117.5 | 1.819297 | 0.000007 | 1.753321 | 0 |
| 1120 | 1.819261 | 0.000007 | 1.753288 | 0 |
| 1122.5 | 1.819226 | 0.000007 | 1.753255 | 0 |
| 1125 | 1.819191 | 0.000007 | 1.753223 | 0 |
| 1127.5 | 1.819156 | 0.000007 | 1.753191 | 0 |
| 1130 | 1.819121 | 0.000007 | 1.753159 | 0 |
| 1132.5 | 1.819087 | 0.000007 | 1.753127 | 0 |
| 1135 | 1.819052 | 0.000007 | 1.753095 | 0 |
| 1137.5 | 1.819018 | 0.000007 | 1.753064 | 0 |
| 1140 | 1.818985 | 0.000007 | 1.753033 | 0 |
| 1142.5 | 1.818951 | 0.000007 | 1.753001 | 0 |
| 1145 | 1.818918 | 0.000007 | 1.752971 | 0 |
| 1147.5 | 1.818884 | 0.000007 | 1.75294 | 0 |
| 1150 | 1.818852 | 0.000007 | 1.752909 | 0 |
| 1152.5 | 1.818819 | 0.000007 | 1.752879 | 0 |
| 1155 | 1.818787 | 0.000007 | 1.752849 | 0 |
| 1157.5 | 1.818754 | 0.000007 | 1.752819 | 0 |
| 1160 | 1.818722 | 0.000007 | 1.752789 | 0 |
| 1162.5 | 1.81869 | 0.000007 | 1.75276 | 0 |
| 1165 | 1.818659 | 0.000007 | 1.75273 | 0 |
| 1167.5 | 1.818627 | 0.000007 | 1.752701 | 0 |
| 1170 | 1.818596 | 0.000007 | 1.752672 | 0 |
| 1172.5 | 1.818565 | 0.000007 | 1.752643 | 0 |
| 1175 | 1.818534 | 0.000007 | 1.752615 | 0 |
| 1177.5 | 1.818504 | 0.000007 | 1.752586 | 0 |
| 1180 | 1.818473 | 0.000007 | 1.752558 | 0 |
| 1182.5 | 1.818443 | 0.000007 | 1.75253 | 0 |
| 1185 | 1.818413 | 0.000007 | 1.752502 | 0 |
| 1187.5 | 1.818383 | 0.000007 | 1.752474 | 0 |
| 1190 | 1.818354 | 0.000007 | 1.752446 | 0 |
| 1192.5 | 1.818324 | 0.000007 | 1.752419 | 0 |
| 1195 | 1.818295 | 0.000007 | 1.752391 | 0 |
| 1197.5 | 1.818266 | 0.000007 | 1.752364 | 0 |
| 1200 | 1.818237 | 0.000007 | 1.752337 | 0 |
| 1202.5 | 1.818209 | 0.000007 | 1.75231 | 0 |
| 1205 | 1.81818 | 0.000007 | 1.752283 | 0 |
| 1207.5 | 1.818152 | 0.000007 | 1.752257 | 0 |
| 1210 | 1.818124 | 0.000007 | 1.75223 | 0 |
| 1212.5 | 1.818096 | 0.000007 | 1.752204 | 0 |
| 1215 | 1.818068 | 0.000007 | 1.752178 | 0 |
| 1217.5 | 1.81804 | 0.000007 | 1.752152 | 0 |
| 1220 | 1.818013 | 0.000007 | 1.752126 | 0 |
| 1222.5 | 1.817985 | 0.000007 | sty.21 | 0 |
| 1225 | 1.817958 | 0.000007 | 1.752075 | 0 |
| 1227.5 | 1.817931 | 0.000007 | 1.752049 | 0 |
| 1230 | 1.817905 | 0.000007 | 1.752024 | 0 |
| 1232.5 | 1.817878 | 0.000007 | 1.751999 | 0 |
| 1235 | 1.817852 | 0.000007 | 1.751974 | 0 |
| 1237.5 | 1.817825 | 0.000007 | 1.751949 | 0 |
| 1240 | 1.817799 | 0.000007 | 1.751925 | 0 |
| 1242.5 | 1.817773 | 0.000007 | sty.19 | 0 |
| 1245 | 1.817747 | 0.000007 | 1.751876 | 0 |
| 1247.5 | 1.817722 | 0.000007 | 1.751851 | 0 |
| 1250 | 1.817696 | 0.000007 | 1.751827 | 0 |
| 1252.5 | 1.817671 | 0.000007 | 1.751803 | 0 |
| 1255 | 1.817646 | 0.000007 | 1.751779 | 0 |
| 1257.5 | 1.817621 | 0.000007 | 1.751756 | 0 |
| 1260 | 1.817596 | 0.000007 | 1.751732 | 0 |
| 1262.5 | 1.817571 | 0.000007 | 1.751708 | 0 |
| 1265 | 1.817546 | 0.000007 | 1.751685 | 0 |
| 1267.5 | 1.817522 | 0.000007 | 1.751662 | 0 |
| 1270 | 1.817498 | 0.000007 | 1.751639 | 0 |
| 1272.5 | 1.817474 | 0.000007 | 1.751616 | 0 |
| 1275 | 1.81745 | 0.000007 | 1.751593 | 0 |
| 1277.5 | 1.817426 | 0.000007 | 1.75157 | 0 |
| 1280 | 1.817402 | 0.000007 | 1.751547 | 0 |
| 1282.5 | 1.817379 | 0.000007 | 1.751525 | 0 |
| 1285 | 1.817355 | 0.000007 | 1.751503 | 0 |
| 1287.5 | 1.817332 | 0.000007 | 1.75148 | 0 |
| 1290 | 1.817309 | 0.000007 | 1.751458 | 0 |
| 1292.5 | 1.817286 | 0.000007 | 1.751436 | 0 |
| 1295 | 1.817263 | 0.000007 | 1.751414 | 0 |
| 1297.5 | 1.81724 | 0.000007 | 1.751392 | 0 |
| 1300 | 1.817217 | 0.000007 | 1.751371 | 0 |
| 1302.5 | 1.817195 | 0.000007 | 1.751349 | 0 |
| 1305 | 1.817173 | 0.000007 | 1.751327 | 0 |
| 1307.5 | 1.81715 | 0.000007 | 1.751306 | 0 |
| 1310 | 1.817128 | 0.000007 | 1.751285 | 0 |
| 1312.5 | 1.817106 | 0.000007 | 1.751264 | 0 |
| 1315 | 1.817084 | 0.000007 | 1.751243 | 0 |
| 1317.5 | 1.817063 | 0.000007 | 1.751222 | 0 |
| 1320 | 1.817041 | 0.000007 | 1.751201 | 0 |
| 1322.5 | 1.81702 | 0.000007 | 1.75118 | 0 |
| 1325 | 1.816998 | 0.000007 | 1.751159 | 0 |
| 1327.5 | 1.816977 | 0.000007 | 1.751139 | 0 |
| 1330 | 1.816956 | 0.000007 | 1.751119 | 0 |
| 1332.5 | 1.816935 | 0.000007 | 1.751098 | 0 |
| 1335 | 1.816914 | 0.000007 | 1.751078 | 0 |
| 1337.5 | 1.816893 | 0.000007 | 1.751058 | 0 |
| 1340 | 1.816873 | 0.000007 | 1.751038 | 0 |
| 1342.5 | 1.816852 | 0.000007 | 1.751018 | 0 |
| 1345 | 1.816832 | 0.000007 | 1.750998 | 0 |
| 1347.5 | 1.816812 | 0.000007 | 1.750978 | 0 |
| 1350 | 1.816791 | 0.000007 | 1.750959 | 0 |
| 1352.5 | 1.816771 | 0.000007 | 1.750939 | 0 |
| 1355 | 1.816751 | 0.000006 | 1.75092 | 0 |
| 1357.5 | 1.816732 | 0.000006 | sty.09 | 0 |
| 1360 | 1.816712 | 0.000006 | 1.750881 | 0 |
| 1362.5 | 1.816692 | 0.000006 | 1.750862 | 0 |
| 1365 | 1.816673 | 0.000006 | 1.750843 | 0 |
| 1367.5 | 1.816653 | 0.000006 | 1.750824 | 0 |
| 1370 | 1.816634 | 0.000006 | 1.750805 | 0 |
| 1372.5 | 1.816615 | 0.000006 | 1.750786 | 0 |
| 1375 | 1.816596 | 0.000006 | 1.750768 | 0 |
| 1377.5 | 1.816577 | 0.000006 | 1.750749 | 0 |
| 1380 | 1.816558 | 0.000006 | 1.75073 | 0 |
| 1382.5 | 1.816539 | 0.000006 | 1.750712 | 0 |
| 1385 | 1.816521 | 0.000006 | 1.750694 | 0 |
| 1387.5 | 1.816502 | 0.000006 | 1.750675 | 0 |
| 1390 | 1.816484 | 0.000006 | 1.750657 | 0 |
| 1392.5 | 1.816465 | 0.000006 | 1.750639 | 0 |
| 1395 | 1.816447 | 0.000006 | 1.750621 | 0 |
| 1397.5 | 1.816429 | 0.000006 | 1.750603 | 0 |
| 1400 | 1.816411 | 0.000006 | 1.750585 | 0 |
| 1402.5 | 1.816393 | 0.000006 | 1.750567 | 0 |
| 1405 | 1.816375 | 0.000006 | 1.75055 | 0 |
| 1407.5 | 1.816357 | 0.000006 | 1.750532 | 0 |
| 1410 | 1.81634 | 0.000006 | 1.750515 | 0 |
| 1412.5 | 1.816322 | 0.000006 | 1.750497 | 0 |
| 1415 | 1.816305 | 0.000006 | 1.75048 | 0 |
| 1417.5 | 1.816287 | 0.000006 | 1.750463 | 0 |
| 1420 | 1.81627 | 0.000006 | 1.750445 | 0 |
| 1422.5 | 1.816253 | 0.000006 | 1.750428 | 0 |
| 1425 | 1.816236 | 0.000006 | 1.750411 | 0 |
| 1427.5 | 1.816218 | 0.000006 | 1.750394 | 0 |
| 1430 | 1.816202 | 0.000006 | 1.750377 | 0 |
| 1432.5 | 1.816185 | 0.000006 | 1.75036 | 0 |
| 1435 | 1.816168 | 0.000006 | 1.750344 | 0 |
| 1437.5 | 1.816151 | 0.000006 | 1.750327 | 0 |
| 1440 | 1.816135 | 0.000006 | 1.75031 | 0 |
| 1442.5 | 1.816118 | 0.000006 | 1.750294 | 0 |
| 1445 | 1.816102 | 0.000006 | 1.750277 | 0 |
| 1447.5 | 1.816086 | 0.000006 | 1.750261 | 0 |
| 1450 | 1.816069 | 0.000006 | 1.750245 | 0 |
| 1452.5 | 1.816053 | 0.000006 | 1.750228 | 0 |
| 1455 | 1.816037 | 0.000006 | 1.750212 | 0 |
| 1457.5 | 1.816021 | 0.000006 | 1.750196 | 0 |
| 1460 | 1.816005 | 0.000006 | 1.75018 | 0 |
| 1462.5 | 1.815989 | 0.000006 | 1.750164 | 0 |
| 1465 | 1.815974 | 0.000006 | 1.750148 | 0 |
| 1467.5 | 1.815958 | 0.000006 | 1.750132 | 0 |
| 1470 | 1.815943 | 0.000006 | 1.750116 | 0 |
| 1472.5 | 1.815927 | 0.000006 | 1.750101 | 0 |
| 1475 | 1.815912 | 0.000006 | 1.750085 | 0 |
| 1477.5 | 1.815896 | 0.000006 | 1.750069 | 0 |
| 1480 | 1.815881 | 0.000006 | 1.750054 | 0 |
| 1482.5 | 1.815866 | 0.000006 | 1.750039 | 0 |
| 1485 | 1.815851 | 0.000006 | 1.750023 | 0 |
| 1487.5 | 1.815836 | 0.000006 | 1.750008 | 0 |
| 1490 | 1.815821 | 0.000006 | 1.749993 | 0 |
| 1492.5 | 1.815806 | 0.000006 | 1.749977 | 0 |
| 1495 | 1.815791 | 0.000006 | 1.749962 | 0 |
| 1497.5 | 1.815777 | 0.000006 | 1.749947 | 0 |
| 1500 | 1.815762 | 0.000006 | 1.749932 | 0 |
| 1502.5 | 1.815747 | 0.000006 | 1.749917 | 0 |
| 1505 | 1.815733 | 0.000006 | 1.749902 | 0 |
| 1507.5 | 1.815718 | 0.000006 | 1.749887 | 0 |
| 1510 | 1.815704 | 0.000006 | 1.749873 | 0 |
| 1512.5 | 1.81569 | 0.000006 | 1.749858 | 0 |
| 1515 | 1.815676 | 0.000006 | 1.749843 | 0 |
| 1517.5 | 1.815661 | 0.000006 | 1.749829 | 0 |
| 1520 | 1.815647 | 0.000006 | 1.749814 | 0 |
| 1522.5 | 1.815633 | 0.000006 | sty.98 | 0 |
| 1525 | 1.815619 | 0.000006 | 1.749785 | 0 |
| 1527.5 | 1.815606 | 0.000006 | 1.749771 | 0 |
| 1530 | 1.815592 | 0.000006 | 1.749757 | 0 |
| 1532.5 | 1.815578 | 0.000006 | 1.749742 | 0 |
| 1535 | 1.815565 | 0.000006 | 1.749728 | 0 |
| 1537.5 | 1.815551 | 0.000006 | 1.749714 | 0 |
| 1540 | 1.815537 | 0.000006 | sty.97 | 0 |
| 1542.5 | 1.815524 | 0.000006 | 1.749686 | 0 |
| 1545 | 1.815511 | 0.000006 | 1.749672 | 0 |
| 1547.5 | 1.815497 | 0.000006 | 1.749658 | 0 |
| 1550 | 1.815484 | 0.000006 | 1.749644 | 0 |
| 1552.5 | 1.815471 | 0.000006 | 1.74963 | 0 |

| 1555 | 1.815458 | 0.000006 | 1.749617 | 0 |
| 1557.5 | 1.815445 | 0.000006 | 1.749603 | 0 |
| 1560 | 1.815432 | 0.000006 | 1.749589 | 0 |
| 1562.5 | 1.815419 | 0.000006 | 1.749576 | 0 |
| 1565 | 1.815406 | 0.000006 | 1.749562 | 0 |
| 1567.5 | 1.815393 | 0.000006 | 1.749549 | 0 |
| 1570 | 1.81538 | 0.000006 | 1.749535 | 0 |
| 1572.5 | 1.815368 | 0.000006 | 1.749522 | 0 |
| 1575 | 1.815355 | 0.000006 | 1.749508 | 0 |
| 1577.5 | 1.815342 | 0.000006 | 1.749495 | 0 |
| 1580 | 1.81533 | 0.000006 | 1.749482 | 0 |
| 1582.5 | 1.815318 | 0.000006 | 1.749469 | 0 |
| 1585 | 1.815305 | 0.000006 | 1.749455 | 0 |
| 1587.5 | 1.815293 | 0.000006 | 1.749442 | 0 |
| 1590 | 1.81528 | 0.000006 | 1.749429 | 0 |
| 1592.5 | 1.815268 | 0.000006 | 1.749416 | 0 |
| 1595 | 1.815256 | 0.000006 | 1.749403 | 0 |
| 1597.5 | 1.815244 | 0.000006 | 1.74939 | 0 |
| 1600 | 1.815232 | 0.000006 | 1.749377 | 0 |
| 1602.5 | 1.81522 | 0.000006 | 1.749365 | 0 |
| 1605 | 1.815208 | 0.000006 | 1.749352 | 0 |
| 1607.5 | 1.815196 | 0.000006 | 1.749339 | 0 |
| 1610 | 1.815184 | 0.000006 | 1.749326 | 0 |
| 1612.5 | 1.815173 | 0.000006 | 1.749314 | 0 |
| 1615 | 1.815161 | 0.000006 | 1.749301 | 0 |
| 1617.5 | 1.815149 | 0.000006 | 1.749289 | 0 |
| 1620 | 1.815138 | 0.000006 | 1.749276 | 0 |
| 1622.5 | 1.815126 | 0.000006 | 1.749264 | 0 |
| 1625 | 1.815115 | 0.000006 | 1.749251 | 0 |
| 1627.5 | 1.815103 | 0.000006 | 1.749239 | 0 |
| 1630 | 1.815092 | 0.000006 | 1.749226 | 0 |
| 1632.5 | 1.815081 | 0.000006 | 1.749214 | 0 |
| 1635 | 1.815069 | 0.000006 | 1.749202 | 0 |
| 1637.5 | 1.815058 | 0.000006 | 1.74919 | 0 |
| 1640 | 1.815047 | 0.000006 | 1.749177 | 0 |
| 1642.5 | 1.815036 | 0.000006 | 1.749165 | 0 |
| 1645 | 1.815025 | 0.000006 | 1.749153 | 0 |
| 1647.5 | 1.815014 | 0.000006 | 1.749141 | 0 |
| 1650 | 1.815003 | 0.000006 | 1.749129 | 0 |
| 1652.5 | 1.814992 | 0.000006 | 1.749117 | 0 |
| 1655 | 1.814981 | 0.000006 | 1.749105 | 0 |
| 1657.5 | 1.81497 | 0.000006 | 1.749093 | 0 |
| 1660 | 1.814959 | 0.000006 | 1.749081 | 0 |
| 1662.5 | 1.814949 | 0.000006 | 1.74907 | 0 |
| 1665 | 1.814938 | 0.000006 | 1.749058 | 0 |
| 1667.5 | 1.814927 | 0.000006 | 1.749046 | 0 |
| 1670 | 1.814917 | 0.000006 | 1.749034 | 0 |
| 1672.5 | 1.814906 | 0.000006 | 1.749023 | 0 |
| 1675 | 1.814896 | 0.000006 | 1.749011 | 0 |
| 1677.5 | 1.814885 | 0.000006 | 1.749 | 0 |
| 1680 | 1.814875 | 0.000006 | 1.748988 | 0 |
| 1682.5 | 1.814865 | 0.000006 | 1.748977 | 0 |
| 1685 | 1.814854 | 0.000006 | 1.748965 | 0 |
| 1687.5 | 1.814844 | 0.000006 | 1.748954 | 0 |
| 1690 | 1.814834 | 0.000006 | 1.748942 | 0 |

**Table S3.** Refractive index of $S_CAlMgO_4$. The values were obtained from the model developed to fit the ellipsometric data.